\documentclass[aps,twocolumn,notitlepage,showpacs,superscriptaddress]{revtex4-1}

\usepackage{cleveref}
\usepackage{graphicx}  
\usepackage{dcolumn}   
\usepackage{bm}        
\usepackage{amssymb}  
\usepackage{amsmath}   
\usepackage{amsthm}
\usepackage{dsfont}
\usepackage[utf8]{inputenc}
\usepackage[english]{babel}
\usepackage{mathtools}
\usepackage{mathrsfs}   
\usepackage{soul}

\usepackage{tikz}
\usetikzlibrary{matrix,arrows.meta,decorations.markings,calc}
\usetikzlibrary{3d}
\usetikzlibrary{decorations.pathreplacing,decorations.markings,calc}

\newcommand{\id}{\mathds{1}}
\newcommand{\tr}{\operatorname{Tr}}
\newcommand{\ket}[1]{\vert #1 \rangle}

\newcommand{\bed}{\[}
\newcommand{\eed}{\]}
\newcommand{\beq}{\begin{equation}}
\newcommand{\eeq}{\end{equation}}

\newcommand{\myinv}[1]{#1^{\scalebox{0.9}[1.0]{-}1}}

\tikzset{
  every picture/.style = {
    baseline={([yshift=-.5ex]current bounding box.center)}, 
    scale=1.2,
    transform shape,
    font=\scriptsize
  }
}

\tikzset{
  3dpeps/.pic={
    \begin{scope}[canvas is zx plane at y=0]
      \draw[thick] (-0.5,0)--(0.5,0);
      \draw[thick] (0,-0.4)--(0,0.4);
      \filldraw (0,0) circle (0.07);
    \end{scope}
    \draw[thick] (0,0,0)--(0,0.2,0);
    
  }
}

\tikzset{
  3dpepsdown/.pic={
    \draw[thick] (0,0,0)--(0,-0.2,0);
    \begin{scope}[canvas is zx plane at y=0]
      \draw[thick] (-0.5,0)--(0.5,0);
      \draw[thick] (0,-0.4)--(0,0.4);
      \filldraw (0,0) circle (0.07);
    \end{scope} 
  }
}

\tikzset{
  3dpepsshort/.pic={
    \begin{scope}[canvas is zx plane at y=0]
      \draw[thick] (-0.5,0)--(0.5,0);
      \draw[thick] (0,-0.3)--(0,0.3);
      \filldraw (0,0) circle (0.07);
    \end{scope}
    \draw[thick] (0,0,0)--(0,0.2,0);
    
  }
}

\tikzset{
  3dpepsdownshort/.pic={
    \draw[thick] (0,0,0)--(0,-0.2,0);
    \begin{scope}[canvas is zx plane at y=0]
      \draw[thick] (-0.5,0)--(0.5,0);
      \draw[thick] (0,-0.3)--(0,0.3);
      \filldraw (0,0) circle (0.07);
    \end{scope} 
  }
}

 \tikzset{
  3dpepsp/.pic={
    \begin{scope}[canvas is zx plane at y=0]
      \draw[thick] (-0.3,0)--(0.3,0);
      \draw[thick] (0,-0.25)--(0,0.25);
      \filldraw[draw=black,fill=blue] (0,0) circle (0.07);
    \end{scope}
  }
}

\tikzset{
  3dpepsdownp/.pic={
    \begin{scope}[canvas is zx plane at y=0]
      \draw[thick] (-0.3,0)--(0.3,0);
      \draw[thick] (0,-0.25)--(0,0.25);
      \filldraw[draw=black,fill=blue] (0,0) circle (0.07);
    \end{scope} 
  }
}
 \tikzset{
  3dpepspshort/.pic={
    \begin{scope}[canvas is zx plane at y=0]
      \draw[thick] (-0.3,0)--(0.3,0);
      \draw[thick] (0,-0.1)--(0,0.1);
      \filldraw[draw=black,fill=blue] (0,0) circle (0.07);
    \end{scope}
  }
}

\tikzset{
  3dpepsdownpshort/.pic={
    \begin{scope}[canvas is zx plane at y=0]
      \draw[thick] (-0.3,0)--(0.3,0);
      \draw[thick] (0,-0.1)--(0,0.1);
      \filldraw[draw=black,fill=blue] (0,0) circle (0.07);
    \end{scope} 
  }
}

 \tikzset{
  3dpepspb/.pic={
    \begin{scope}[canvas is zx plane at y=0]
      \draw[thick] (-0.3,0)--(0.3,0);
      \draw[thick] (0,-0.25)--(0,0.25);
      \filldraw[draw=black,fill=black] (0,0) circle (0.07);
    \end{scope}
  }
}

\tikzset{
  3dpepsdownpb/.pic={
    \begin{scope}[canvas is zx plane at y=0]
      \draw[thick] (-0.3,0)--(0.3,0);
      \draw[thick] (0,-0.25)--(0,0.25);
      \filldraw[draw=black,fill=black] (0,0) circle (0.07);
    \end{scope} 
  }
}

\tikzset{
  3dGisopeps/.pic={
       \filldraw (-0.3,0,0) circle (0.04);
     \filldraw (0,0,0.3) circle (0.04);
       \filldraw (0.3,0,0) circle (0.04);
        \filldraw (0,0,-0.3) circle (0.04); 
\draw (-0.1,0,0)--(-0.1,0.4,0);
       \begin{scope}[canvas is zx plane at y=0]
          \draw[thick] (0.1,0)--(0.5,0);
        \draw[thick] (-0.5,0)--(-0.1,0);
    \end{scope} 
     \draw[thick] (0,0,0.1)--(0,0.4,0.1);
\draw[thick] (0,0,-0.1)--(0,0.4,-0.1);
       \draw[preaction={draw, line width=1pt, white}] (0.1,0,0)--(0.1,0.4,0);
         \begin{scope}[canvas is zx plane at y=0]
          \draw[thick] (0,0.1)--(0,0.4);
       \draw[thick] (0,-0.4)--(0,-0.1);
          \end{scope} 
  }
}

\tikzset{
  3dGisopepsproj/.pic={
         \filldraw (-0.3,0,0) circle (0.04);
     \filldraw (0,0,0.3) circle (0.04);
       \filldraw (0.3,0,0) circle (0.04);
        \filldraw (0,0,-0.3) circle (0.04); 
\draw (-0.1,0,0)--(-0.1,0.4,0);
       \begin{scope}[canvas is zx plane at y=0]
          \draw (0.1,0)--(0.5,0);
        \draw (-0.5,0)--(-0.1,0);
    \end{scope} 
     \begin{scope}[canvas is zx plane at y=0.4]
       \draw (-0.5,0)--(-0.1,0);
               \draw [->] (-0.1,0)--(-0.4,0);
      \draw[preaction={draw, line width=1pt, white}] (0.1,0)--(0.5,0);
      \draw [->,preaction={draw, line width=1pt, white}]  (0.5,0)--(0.2,0);
    \end{scope} 
     \draw (0,0,0.1)--(0,0.4,0.1);
\draw (0,0,-0.1)--(0,0.4,-0.1);
       \draw[preaction={draw, line width=1pt, white}] (0.1,0,0)--(0.1,0.4,0);
         \begin{scope}[canvas is zx plane at y=0]
          \draw (0,0.1)--(0,0.4);
       \draw (0,-0.4)--(0,-0.1);
          \end{scope} 
                  \begin{scope}[canvas is zx plane at y=0.4]
          \draw (0,0.1)--(0,0.4);
           \draw [->] (0,0.1)--(0,0.3);
       \draw (0,-0.4)--(0,-0.1);
        \draw [->] (0,-0.3)--(0,-0.2);
          \end{scope} 
  }
}

\renewcommand{\thefootnote}{\fnsymbol{footnote}}

\begin{document}

\title{Local order parameters for symmetry fractionalization}

\author{Jos\'e Garre-Rubio}
\email[Corresponding author: ]{jogarre@ucm.es}
\affiliation{Departamento de An\'alisis  y Matem\'atica Aplicada, Universidad Complutense de Madrid, Spain \\
Instituto de Ciencias Matem\'aticas, Madrid, Spain}
\author{Sofyan Iblisdir}
\affiliation{Dpt. de F\'isica Qu\`antica i Astronomia \& Institut de Ci\`encies del Cosmos, Facultat de F\'isica, Universitat de Barcelona, Barcelona, Spain}

\renewcommand{\thefootnote}{\arabic{footnote}}
\begin{abstract}

We propose a family of order parameters to detect the symmetry fractionalization class of anyons in 2D topological phases. 
This fractionalization class accounts for the \emph{projective}, as opposed to linear, representations of the symmetry group on the anyons.
We focus on quantum double models on a lattice enriched with an internal symmetry in the framework of $G$-isometric projected entangled pair states.
Unlike previous schemes based on reductions to effective 1D systems (dimensional compactification), the order parameters presented here can be probed on genuinely two-dimensional geometries, and are \emph{local}: they rely on operations on few neighbouring particles in the bulk.
The power of these order parameters is illustrated with several combinations of topological content and symmetry. We demonstrate that a strictly finer phase distinction than that provided by dimensional compactification can be obtained. As particular examples, the resolution power of these order parameters is illustrated for a case with non-abelian topological order, and for another with symmetries that involves permutation of anyons.
\end{abstract}

\maketitle

\section{Introduction}

One of the cornerstones of condensed matter physics is the study of quantum phases, that is, the different features that zero temperature systems can exhibit. The phases are the classes that result from deeming as equivalent states that can be connected via adiabatic paths \cite{Hastings05,Bravyi10,Chen10}.  In two spatial dimensions, even in the absence of symmetry restrictions regarding such paths, there are quantum systems that cannot be adiabatically transformed into each other. These states possess different topological orders, \emph{i.e.} different patterns of long range entanglement \cite{Wen04}. One of the most important features of topological phases is the existence of quasiparticle excitations (anyons) which exhibit unusual exchange statistics \cite{Wilckez82, Nayak08}. When the  adiabatic path between pairs of states is required to preserve some symmetry, the phase diagram becomes richer. There have been a tremendous effort to classify these phases in the last decade, see \emph{e.g.} \cite{Chen13, Essin13, Barkeshli14} and references therein. 

In absence of topological order, the phases that appear by preserving a symmetry are called Symmetry Protected Topological ({SPT}) phases. A well-known example is the Haldane phase \cite{Haldane93A, Haldane93B}. When there is non-trivial topological order, the classes obtained are denoted as Symmetry Enriched Topological ({SET}) phases \cite{Barkeshli14}. When a global symmetry on the quantum state of an {SET} phase is applied, although this state is left invariant, its excitations need not be. They may transform projectively; we then say that the global symmetry is fractionalized. An important example of such a scenario is the fractional quantum Hall effect, where for the filling factor $\nu=1/3$ the anyons carry a third of the electron charge \cite{Laughlin83,Tsui82}. The different patterns of Symmetry Fractionalization ({SF}) give rise to inequivalent quantum phases. Their identification therefore plays a fundamental role when studying {SET} phases.
 
In this work, we propose order parameters to identify the {SF} classes of the topological quasiparticles. To do so, we will exploit an equivalence between symmetry fractionalization of an anyon and braiding this anyon with another associated anyon \cite{Barkeshli14,Chen17}. The type of this associated anyon determines fully the {SF} class, \emph{i.e.} the {SET} phase.
Previous works \cite{Cincio15,Zaletel16, Zaletel17, Wang15, Qi15, Saadarmand16, Huang14} use one-dimensional compactification to study the {SF} of the anyons. The studied {2D} system is given the geometry of a long cylinder, and one-dimensional techniques, used in the context of {SPT} phases, are employed to detect the {SF} class of the {2D} system. Refs. \cite{Cincio15,Zaletel16, Zaletel17, Wang15, Qi15, Saadarmand16} focus on lattice and time reversal symmetries, on-site symmetries are dealt with in \cite{Huang14}. 
Here, we will introduce a family of order parameters that is genuinely {2D}; we do not use a one-dimensional compactification which could miss intrinsic {2D} features or prevent physical implementations. These order parameters are \emph{local}: their supports are finite regions of the lattice. 
They are similar in spirit to that proposed in \cite{Haegeman12}  for {1D} {SPT} phases.

\

We focus on quantum double models of a group $G$, $\mathcal{D}(G)$\cite{Schuch10}, the generalization of the toric code model \cite{Kitaev03}, enriched with an on-site symmetry  of a finite group, $Q$ \cite{Jiang15,Garre17,Williamson17}. 
We will consider the pairs $( G, Q ) = ( \mathbb{Z}_2, \mathbb{Z}_2 ), ( \mathbb{Z}_2, \mathbb{Z}_2 \times \mathbb{Z}_2 ), \{ (\mathbb{Z}_p, \mathbb{Z}_p): p \; \text{prime} \}, ( \mathbb{Z}_4, \mathbb{Z}_2 ), ( Q_8, \mathbb{Z}_2 )$ and exhibit order parameters that allow to fully resolve between the various {SET} phases. These instances have been chosen because they illustrate simply a variety of scenarios, and why {SET} detection based on dimensional compactification may miss {SF} patterns. For all these examples, a \emph{strictly finer} phase resolution will be demonstrated. In the simplest case of the toric code, $( \mathbb{Z}_2, \mathbb{Z}_2 )$, we will find two distinct {SF} classes for which the {SPT} order parameter assumes the same value. The case $( \mathbb{Z}_2, \mathbb{Z}_2 \times \mathbb{Z}_2 )$ is interesting in that the {SPT} order parameter exhibits \emph{some coarse} distinction. The case $(\mathbb{Z}_4, \mathbb{Z}_2)$ involves permutation of anyon types; the case $( Q_8, \mathbb{Z}_2 )$ is particular in that the topological content is non-abelian. 

Our work is heavily based on the formalism of tensor network states. This framework has turned out to be adequate at accurately and systematically describing quantum phases. In {1D}, Matrix Product States ({MPS}) have been used successfully both numerically \cite{White92A,Schollwock11} and formally \cite{Hastings07A,Hastings07B} to capture the behaviour of ground states of local gapped Hamiltonians. Within this framework, the classification of {SPT} phases in {1D} has been achieved \cite{Chen11,Schuch11}. The analogous {2D} tensor networks, the so-called Projected Entangled Pair States (PEPS), have also proven to be powerful tools. For example, it has been shown that they describe the renormalization fixed point of all known non-chiral topological phases in bosonic {2D} systems without any symmetry \cite{Schuch10, Sahinoglu14} as well as those enriched with symmetries \cite{Jiang15, Garre17, Williamson17}. 

The paper is organized as follows. Section II is a reminder on {PEPS} and some properties of $G$-injective {PEPS}. The reader familiar with these notions can skip it altogether. In Section III, we describe how symmetries are realised in $G$-injective {PEPS}, and we present our order parameters in their renormalization fixed points called $G$-isometric PEPS. Section IV is devoted to the examples metioned above. We conclude and discuss open questions in Section V.

\section{{PEPS}, topological order and excitations}

\begin{figure}[ht!]
\begin{center}
\includegraphics[scale=1.2]{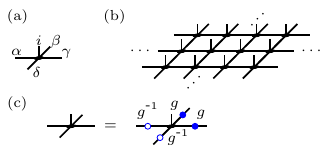}
\caption{(a) Diagrammatic picture of the tensor $A^i_{\alpha \beta \gamma \delta}$ of a {PEPS} for the square lattice. (b) The {PEPS} constructed via the concatenation of the virtual d.o.f. of the tensors (without boundary conditions). Each leg corresponds to an index, index contraction is represented by joining legs. We will always assume periodic boundary condition; the tensors are placed on a torus. We will however leave the boundaries open in the drawings for the sake of clarity. (c) Invariance of the $G$-injective tensor under the action of the group $G$  on the virtual d.o.f. ($g$ stands for $u_g$.)}
\label{fig:PEPSconstruction}
\end{center}
\end{figure}

{PEPS} are pure states on a graph of identical (bosonic) particles fully characterised by a tensor $A$ associated with each vertex (for a thorough introduction see \cite{Orus14} and references therein). We focus on translational invariant systems placed on a square lattice. The tensor $A^i_{\alpha,\beta,\gamma,\delta}$ -see Fig.\ref{fig:PEPSconstruction}(a)- then has five indices; one $i=1,\dots,d$ describing the physical Hilbert space $\mathbb{C}^d$ of each particle and four $\alpha,\beta,\gamma,\delta= 1,\dots, D$ corresponding to the virtual degrees of freedom (d.o.f.). The {PEPS} is constructed by associating a copy of $A$ with each vertex and contracting the neighbour virtual d.o.f. (identifying and summing the indices) as depicted in Fig.\ref{fig:PEPSconstruction}(b). When the chosen boundary conditions are applied, the resulting tensor contraction $\psi_{i_1,\cdots,i_N}=\mathcal{C}\{ A^{i_1},\dots, A^{i_N}\}$ describes a quantum many-body state $|\psi(A)\rangle =\sum \psi_{i_1,\cdots, i_N}|i_1\cdots i_N\rangle$. 

The family of {PEPS} we are interested in has the following symmetry illustrated on Fig.\ref{fig:PEPSconstruction}(c):
$$A=A(u_g \otimes u_g \otimes \myinv{u}_g \otimes \myinv{u}_g ),$$
where $u_g$ is some unitary representation of the finite group $G$ acting on the virtual d.o.f. of the {PEPS}. This class of {PEPS}, termed $G$-injective \cite{Schuch10}, are such that there exists some tensor, which we will denote $\myinv{A}$, that satisfies:
\begin{equation}\label{Ginjec}
\begin{tikzpicture}
   \pic[thick] at (0,0,0) {3dpepsshort};
    \pic[thick] at (0,0.4,0) {3dpepsdownshort};
    \node[anchor=north] at (0,0,0) {${A}$};
    \node[anchor=south] at (0,0.4,0) {$\myinv{A}$};
 \end{tikzpicture}
=\frac{1}{|G|}\sum_{g\in G} u_g \otimes u_g \otimes \myinv{u}_g \otimes \myinv{u}_g 
 =  \frac{1}{|G|}\sum_{g\in G}  
     \begin{tikzpicture}
       \pic[thick] at (0,0,0) {3dGisopepsproj};
                      \node at (-0.25,0.2,0.3) {$\myinv{g}$};
              \node at (0.15,-0.1,0.3) {$\myinv{g}$};
        \node at (0.37,0.35,0.3) {${g}$};
                \node at (0.5,-0.03,0.3) {${g}$};
\end{tikzpicture},
\end{equation}
where we have written $g$ as a shorthand for $u_g$ (we will do it through the manuscript) and the arrows establish the orientation that we will follow through the whole manuscript.

Given a $G$-injective {PEPS}, the associated parent Hamiltonian \cite{Schuch10}, defined on a finite torus, has the ground state degeneracy of the topological model corresponding to $\mathcal{D}$($G$). An important subfamily are the so-called $G$-isometric {PEPS} in which  $u_g=L_g$ is the left regular representation, acting as $L_g|h\rangle=|gh\rangle$ on $\mathbb{C}[G]={\rm span}\{|h\rangle, h\in G\}$. 
In this case $A^{-1}=\bar{A}$.  Using $\tr[L_g]=|G|\delta_{e,g}$ for all $g\in G$, it follows that the contraction of two neighbouring sites reads:
\begin{equation}\label{Gisoconca}
\begin{tikzpicture}
   \pic[thick] at (0,0,0) {3dpeps};
    \pic[thick] at (0,0.4,0) {3dpepsdown};
    \node[anchor=north] at (0,0,0) {${A}$};
    \node[anchor=south] at (0,0.4,0) {$\bar{A}$};
       \pic[thick] at (0.6,0,0) {3dpeps};
    \pic[thick] at (0.6,0.4,0) {3dpepsdown};
    \node[anchor=north] at (0.6,0,0) {${A}$};
    \node[anchor=south] at (0.6,0.4,0) {$\bar{A}$};
  \end{tikzpicture}
   =  \frac{1}{|G|}\sum_{g\in G} \;  
\begin{tikzpicture}
         \filldraw (-0.3,0,0) circle (0.04);
     \filldraw (0,0,0.3) circle (0.04);
       \filldraw (0.8,0,0) circle (0.04);
        \filldraw (0,0,-0.3) circle (0.04); 
             \filldraw (0.5,0,0.3) circle (0.04);
        \filldraw (0.5,0,-0.3) circle (0.04); 
                            \node at (-0.25,0.2,0.3) {$\myinv{g}$};
              \node at (0.15,-0.1,0.3) {$\myinv{g}$};
                \node at (0.65,-0.1,0.3) {$\myinv{g}$};
        \node at (0.25,0.37,0.3) {${g}$};
                \node at (1,0,0.3) {${g}$};
                     \node at (0.9,0.33,0.3) {${g}$};
\draw[thick] (-0.1,0,0)--(-0.1,0.4,0);
       \begin{scope}[canvas is zx plane at y=0]
          \draw[thick] (0.1,0)--(0.5,0);
        \draw[thick] (-0.5,0)--(-0.1,0);
           \draw[thick] (0.1,0.5)--(0.5,0.5);
        \draw[thick] (-0.5,0.5)--(-0.1,0.5);
    \end{scope} 
     \begin{scope}[canvas is zx plane at y=0.4]
       \draw[thick] (-0.5,0)--(-0.1,0);
        \draw[->] (-0.1,0)--(-0.4,0);
      \draw[thick, preaction={draw, line width=1pt, white}] (0.1,0)--(0.5,0);
      \draw[->] (0.5,0)--(0.2,0);
             \draw[thick] (-0.5,0.5)--(-0.1,0.5);
             \draw[->] (-0.1,0.5)--(-0.4,0.5);
      \draw[thick, preaction={draw, line width=1pt, white}] (0.1,0.5)--(0.5,0.5);
                   \draw[->] (0.5,0.5)--(0.2,0.5);
    \end{scope} 
     \draw[thick] (0,0,0.1)--(0,0.4,0.1);
\draw[thick] (0,0,-0.1)--(0,0.4,-0.1);
     \draw[thick] (0.5,0,0.1)--(0.5,0.4,0.1);
\draw[thick] (0.5,0,-0.1)--(0.5,0.4,-0.1);
       \draw[thick,preaction={draw, line width=1pt, white}] (0.6,0,0)--(0.6,0.4,0);
         \begin{scope}[canvas is zx plane at y=0]
          \draw[thick] (0,0.6)--(0,0.9);
       \draw[thick] (0,-0.4)--(0,-0.1);
          \end{scope} 
                  \begin{scope}[canvas is zx plane at y=0.4]
          \draw[thick] (0,0.6)--(0,0.9);
          \draw[->] (0,0.6)--(0,0.8);
       \draw[thick] (0,-0.4)--(0,-0.1);
        \draw[->] (0,-0.4)--(0,-0.2);
          \end{scope}      
\end{tikzpicture}.
 \end{equation}
 Eq.\eqref{Gisoconca} allows us to compute easily expectation values since $\langle\psi(A) |=|\psi(A)\rangle ^\dagger=\sum \mathcal{C}\{ \bar{A}^{i_1},\dots, \bar{A}^{i_N}\} \langle i_1\cdots i_N|$.
 Concretely, considering a connected region $\mathcal{M}$, the contraction of each tensor $A$ inside $\mathcal{M}$ with its corresponding $\bar{A}$ tensor results in the boundary operator $\frac{1}{|G|}\sum_{b\in G} L_b\otimes \cdots \otimes L^{-1}_b$, where each term of the sum contains $|\partial \mathcal{M}|$ factors. 
 
The expectation value of a local operator acting on the complementary of $\mathcal{M}$, $\mathcal{M}^{c}$, is computed by applying the previous boundary operator to the contraction between the tensors $A$ in $\mathcal{M}^{c}$, the local operator and the $\bar{A}$ in $\mathcal{M}^{c}$.
 

$G$-isometric {PEPS} are renormalization group fixed points (RFP) in the phase of the $\mathcal{D}$($G$) \cite{Schuch10}; they are characterised by a commuting parent Hamiltonian supporting anyonic excitations. There are two kinds of anyons of the $\mathcal{D}$($G$) relevant to our work: charges and fluxes characterized by the irreducible representations (irreps) and the conjugacy classes of $G$ respectively. These excitations are locally created in particle-antiparticle pairs that can be moved around unitarily \cite{Schuch10}.
We will denote by $\mathcal{O}_{ \sigma} (x,y)$  the physical operator that creates a particle-antiparticle of type $\sigma$ in the edges $x,y$ respectively, where $\sigma$ is an irrep of $G$.

The creation of such excitations can be expressed through operators acting on the virtual d.o.f. of the {PEPS}. For convenience, we will always work with such operators. The virtual operator associated with a pair of charges is $\Pi_{\sigma}=\sum_{g,h\in G} \chi_\sigma(h^{-1}g)|g\rangle \langle g|\otimes |h\rangle \langle h|$ where $\chi_\sigma$ is the character of $\sigma$ and the tensor product reflects the fact that the two particles are acting on two different virtual edges. If $\sigma$ is a one-dimensional irrep, $\Pi_{\sigma}$ can be factorised: $\Pi_{\sigma}=\left(\sum_{g\in G} \chi_\sigma(g)|g\rangle \langle g|\right)\otimes \left(\sum_{g\in G} \chi_\sigma(h^{-1}) |h\rangle \langle h|\right) \equiv C_{\sigma}\otimes \bar{C}_{\sigma}$. Else, $\Pi_\sigma$ is a sum of factors: $\Pi_{\sigma}\equiv\sum_{h\in G} C_{\sigma,h}\otimes \bar{C}_{\sigma,h}$. In any event, $\Pi_{\sigma}$ will be represented as two orange rectangles on two different edges of the lattice: 
 \begin{equation}      
 \begin{tikzpicture}[scale=1.2]
\foreach \x in {0,0.5,1,1.5}{
	\foreach \z in {0,0.7,1.4}{
  	 \pic[thick] at (\x,0,\z) {3dpeps};}}
     	\filldraw[thick, draw=black,fill=orange] (0.25,0,0.55) rectangle (0.25,0,0.85);           
	\filldraw[thick, draw=black,fill=orange] (1.25,0,0.55) rectangle (1.25,0,0.85);
 \end{tikzpicture} 
. \notag
\end{equation}

\section{Order parameters for symmetry fractionalization}

\subsection{Symmetries of $\mathbf{G}$-injective {PEPS}}

We consider $G$-injective {PEPS},  $|\psi(A)\rangle$, with periodic boundary conditions. $|\psi(A)\rangle$ can be thought of as the vacuum background of some Hamiltonian with topological content characterised by $\mathcal{D}(G)$. We further assume a global on-site symmetry $U_q^{\otimes n}|\psi (A)\rangle=|\psi (A)\rangle \;\forall q\in Q,$ where $U_q$ is a linear unitary representation of some finite group $Q$ and $n$ is the number of lattice sites. 
For all $ q \in Q $, there is an invertible matrix $v_q$ which translates $U_q$ through the local tensor $A$ on the virtual d.o.f. as follows \cite{Molnarinprep}: 
 
  \begin{equation} \label{locsym}
    \begin{tikzpicture}[baseline=-1mm, scale=1.2]
      \pic at (0,0,0) {3dpeps};
      \draw[thick] (0,0,0) -- (0,0.35,0);
      \filldraw[draw=black,fill=purple,thick] (0,0.2,0) circle (0.07);
      \node[anchor=east] at (0,0.3,0) {$U_q$};
    \end{tikzpicture} =
    \begin{tikzpicture}[scale=1.2]
      \draw[thick] (-0.7,0,0) -- (0.7,0,0);
      \draw[thick] (0,0,-0.9) -- (0,0,0.9);
      \pic at (0,0,0) {3dpeps};
      \filldraw[draw=black,fill=red,thick] (-0.5,0,0) circle (0.06);
      \filldraw[draw=black,fill=red,thick] (0.5,0,0) circle (0.06);
      \node[anchor=south] at (-0.5,0,0) {$\myinv{v_q}$};
      \node[anchor=north] at (0.5,0,0) {$v_q$};
      \filldraw[draw=black,fill=red,thick] (0,0,-0.6) circle (0.06);
      \filldraw[draw=black,fill=red,thick] (0,0,0.6) circle (0.06);
       \node[anchor=south] at (0,0,-0.6) {$v_q$};
      \node[anchor=north] at (0,0,0.6) {$\myinv{v_q}$};
    \end{tikzpicture}
    \; \forall q\in Q.
  \end{equation}
  
The operators  $v_q$ do not have to form a linear representation. It actually turns out that 
$$ v_k v_q = {\omega(k,q)} \; v_{kq},$$
where  $G \ni  \omega(k,q) \neq 1$, \emph{i.e.} $v_q$ is a \emph{projective} representation (see Appendix \ref{projrep}) . 
$\{ v_q \}$ may fail to form a homomorphism by a matrix: $\omega(k,q)\equiv u_{\omega(k,q)}$, where $u$ is the representation of $G$ acting on the virtual d.o.f. introduced earlier. Since the representation $\{U_q : q \in Q \}$ is assumed to be linear, the projective nature of $v_q$ does not show up in the action over $\ket{\psi(A)}$. Indeed, Eq.(\ref{locsym}) shows that the operators $v_q$ cancel in pairs on each virtual bond. But the situation changes in regions that contain quasiparticle excitations. For example, one can use  Eq.(\ref{locsym}) to calculate how the on-site symmetry $U_q$ affects a charge sitting on a virtual bond: $C_{\sigma,h} \to \Phi_q(C_{\sigma,h})$, where $\Phi:Q\to \mathcal{L}(\mathcal{M}_D): q \to \Phi_q$, $\Phi_q : X \to \Phi_q( X ) = v_q X v^{-1}_q$.

Diagramatically: 
\begin{equation*}
  \begin{tikzpicture}[scale=1.2]
   \pic at (0,0,0.7) {3dpepsshort}; 
        \pic at (0,0,2.1) {3dpepsshort};
        \pic at (0,0,1.4) {3dpepsshort}; 
      \pic at (0.5,0,0.7) {3dpepsshort};
      \pic at (0.5,0,1.4) {3dpepsshort};
     \pic at (0.5,0,2.1) {3dpepsshort};
     	 \filldraw[draw=black,fill=purple,thick] (0,0.11,0.7) circle (0.05);
	 \filldraw[draw=black,fill=purple,thick] (0,0.11,2.1) circle (0.05);
	 \filldraw[draw=black,fill=purple,thick] (0,0.11,1.4) circle (0.05);
	 \filldraw[draw=black,fill=purple,thick] (0.5,0.11,0.7) circle (0.05);
	 \filldraw[draw=black,fill=purple,thick] (0.5,0.11,1.4) circle (0.05);
	 \filldraw[draw=black,fill=purple,thick] (0.5,0.11,2.1) circle (0.05);
	    \node[anchor=south] at (-0.1,0.15,1)  {$U_q$};
	         \filldraw[draw=black,fill=orange,thick] (0.15,0,1.27) rectangle (0.35,0,1.53); 
 \end{tikzpicture} 
 =
 \begin{tikzpicture}[scale=1.2]
     \pic at (0,0,0.7) {3dpepsshort}; 
        \pic at (0,0,2.1) {3dpepsshort};
        \pic at (0,0,1.4) {3dpepsshort}; 
      \pic at (0.5,0,0.7) {3dpepsshort};
      \pic at (0.5,0,1.4) {3dpepsshort};
     \pic at (0.5,0,2.1) {3dpepsshort};

         \filldraw[draw=black,fill=red,thick]  (0.12,0,1.4) circle (0.04);
	  \filldraw[draw=black,fill=orange,thick] (0.15,0,1.27) rectangle (0.35,0,1.53); 
	\filldraw[draw=black,fill=red,thick]  (0.38,0,1.4) circle (0.04);
 \end{tikzpicture} 
,
  \begin{tikzpicture}[scale=1.2]
      \draw[thick] (0.1,0,0) -- (0.7,0,0);
      \filldraw[draw=black,fill=red,thick] (0.2,0,0) circle (0.05);
      \filldraw[draw=black,fill=red,thick] (0.6,0,0) circle (0.05);
      \node[anchor=south] at (0.2,0,0) {$v_q$};
      \node[anchor=north] at (0.65,0,-0.15) {$\myinv{v_q}$};
       \filldraw[draw=black,fill=orange,thick] (0.4,0,-0.2) rectangle (0.4,0,0.2); 
    \end{tikzpicture}
        \equiv  \Phi_q(C_{\sigma,h}).
 \end{equation*}
If the symmetry is applied for two elements $q,k\in Q$, we see that 
\begin{equation}\label{eq:symcharge}
(\Phi_k\circ \Phi_q)(C_{\sigma,h})=(\tau_{\omega(k,q)}\circ \Phi_{kq})(C_{\sigma,h}),
\end{equation}
where $\tau_\omega$ denotes conjugation by $\omega \in G$. This implies that the symmetry action over the charge sector can be \emph{projective}, \emph{i.e.} the symmetry fractionalizes. This possibility exists because only states with total zero charge are required to transform linearly under the \emph{global} symmetry. 
When a symmetry acts locally around each anyon, projective representations are allowed as long as they combine so that the global symmetry is linear. For example in abelian theories, given an anyon $\sigma$, its anti-particle $\bar{\sigma}$ transforms as the inverse projective representation (picking up the conjugate phase factor) \cite{Chen17}. 

If we knew the matrices $v_q$, we could identify $\omega$. But in general, this is impossible. It is our goal to construct order parameters that identify the projective representation of the charges without relying on the knowledge of $v_q$.\\

In the following we will analyze the properties of the virtual symmetry operators in more details. Two maps constructed with the operators $\{v_q\}$ characterize the interplay between symmetry and topology \cite{Molnarinprep}, the so-called cocycle and twisting map \cite{Adem04}: 
\begin{align*}
\omega : Q \times Q \to G: (k,q) \to \omega(k,q) = v_k v_q v^{-1}_{kq},\\ 
\phi: Q \to \textrm{Aut}( G ): q \to \phi_q, \end{align*}
where $\phi_q( g ) = v_q g v_q^{-1} \; \forall g \in G$. Associativity of group multiplication implies the so-called cocycle condition
\begin{equation*}\label{eq:2cocyclecond}
\omega(k,q) \omega(kq,p)= \phi_k ( \omega(q,p))  \omega(k,qp).
\end{equation*}
These two maps are not independent:
\begin{equation*}
v_k v_q= \omega(k,q)v_{kq}\Rightarrow \phi_k\circ \phi_q=\tau_{\omega(k,q)}\circ \phi_{kq}.
\end{equation*}
 $\omega \neq 1$ provides situations where $\phi$ is \emph{not}  a homomorphism from $Q$ to Aut($G$);  \emph{i. e. } the operators $v_q$ constitute a projective representation of $Q$. The operator $v_q$ is defined up to an element of $G$, for each $q\in Q$, because of the $G$-injectivity of the tensor (see Eq. \eqref{locsym}). Therefore, the equivalence $v_q\sim lv_q$, where $l\in G$, has to be considered as a gauge freedom. The gauge freedom of the virtual operators modifies $\omega$ and $\phi$ as follows: 
\begin{align}\label{cocylefreedom}
\omega'(k,q)&=\ l_kv_kl_q v_q v^{-1}_{kq}l^{-1}_{kq}=l_k \phi_k(l_q)\omega(k,q)l^{-1}_{kq}, \notag \\
\phi'_q &= \tau_{l_q} \circ \phi_q. 
\end{align}
Thus, two pairs cocycle/twisting map, $(\phi,\omega)$ and $(\phi',\omega')$, related by Eq.(\ref{cocylefreedom}) have to be considered equivalent. For fixed $\phi$, the possible cocycles coming from $G$ and $Q$ are classified according to the above equivalence relation and it turns out \cite{Molnarinprep} that they form discrete labels which characterise the different symmetric phases of $G$-injective {PEPS}. When $G$ is abelian, the cocycles are classified by the second cohomology group $H^2_\phi(Q,G)$ \cite{Adem04}. When $G$ is non-abelian the classification is more subtle, as will be seen in Section \ref{ex:nonab}.
 
The pair $(\phi,\omega)$ characterises the action of the symmetry over the anyons, when considering the gapped parent Hamiltonian of $G$-isometric {PEPS}. If $\phi$ or $\omega$ is non-trivial, we are in a non-trivial {SET} phase. Concretely, $\phi$ describes the permutation of the anyon types (and ground states on the torus) whereas $\omega$ describes the {SF} effect on charges (see \cite{Barkeshli14} for the general picture). The \emph{projective} actions associated with the anyons have to be compatible with the fusion rules of the theory and it turns out that they are equivalent to the braiding with some anyon \cite{Barkeshli14}(that characterizes the {SF} class). In the following subsections we show how this relation particularises for {PEPS}.

\subsection{Braiding in $G$-injective {PEPS}}

Let us study braiding in $G$-isometric {PEPS}, and consider a process where a charge -- anti-charge pair $\Pi_{\sigma}$ is created and a flux, characterised by the conjugacy class $[g]$, is braided counter-clockwise around one of them.
The creation of a pair of fluxes associated with some group element $g$, is described by a tensor product of operators, $\bigotimes_{ i \in \gamma } (L_g)^{m_i}$, where $\gamma$ denotes a path of lattice edges connecting the plaquettes where each flux is located. For each link $i$, $m_i \in \{ -1, + 1 \}$; the precise value of $m_i$ depends on the path $\gamma$, see Ref.\cite{Schuch10} for details. Diagrammatically, we will represent such operators as a string of blue circles or circumferences on the edges of the square lattice. The excitations are placed at the end plaquettes. Using Fig. \ref{fig:PEPSconstruction}(c), we see that braiding one flux around a charged link results in the conjugation by $L^\dagger_g$ of the charge operator:

	 \begin{equation*}      
	 \begin{tikzpicture}[scale=1.2]
        \pic at (0,0,0.7) {3dpepsshort};
      \pic at (0.5,0,0) {3dpeps};
      \pic at (0.5,0,0.7) {3dpeps};
      \pic at (0.5,0,1.4) {3dpeps};
     \pic at (1,0,0) {3dpeps};
        \pic at (1,0,0.7) {3dpeps};
        \pic at (1,0,1.4) {3dpeps};
        \pic at (1.5,0,0.7) {3dpeps};
  \pic at (2,0,0.7) {3dpepsshort};
     	\filldraw[draw=black,fill=orange,thick] (0.75,0,0.55) rectangle (0.75,0,0.85);           
	\filldraw[draw=black,fill=orange,thick] (1.75,0,0.55) rectangle (1.75,0,0.85);
		\filldraw[blue,thick]  (0.75,0,1.4) circle (0.04);
		 \draw[densely dotted, blue, rounded corners,thick] (0.75,0,2)--(0.75,0,1.05) --(1.25,0,1.05)--(1.25,0,0.35)-- (0.25,0,0.35)--(0.25,0,1.05)--(0.5,0,1.05);
		 \filldraw[draw=blue,fill=white,thick]  (1,0,1.05) circle (0.04);
		 \filldraw[blue]  (1.25,0,0.7) circle (0.04);
		 \filldraw[blue]  (1,0,0.35) circle (0.04);
		  \filldraw[draw=blue,fill=white,thick]  (0.5,0,1.05) circle (0.04);
		 \filldraw[draw=blue,fill=white,thick]  (0.25,0,0.7) circle (0.04);
		 \filldraw[blue]  (0.5,0,0.35) circle (0.04);
	\end{tikzpicture}
	=
	 \begin{tikzpicture}[scale=1.2]
        \pic at (0,0,0.7) {3dpepsshort};
      \pic at (0.5,0,0) {3dpeps};
      \pic at (0.5,0,0.7) {3dpeps};
      \pic at (0.5,0,1.4) {3dpeps};
     \pic at (1,0,0) {3dpeps};
        \pic at (1,0,0.7) {3dpeps};
        \pic at (1,0,1.4) {3dpeps};
        \pic at (1.5,0,0.7) {3dpeps};
  \pic at (2,0,0.7) {3dpepsshort};
  \filldraw[draw=blue,fill=white,thick]  (0.63,0,0.7) circle (0.04);
     	\filldraw[draw=black,fill=orange,thick] (0.75,0,0.55) rectangle (0.75,0,0.85);   
	  \filldraw[blue,thick]  (0.87,0,0.7) circle (0.04);        
	\filldraw[draw=black,fill=orange,thick] (1.75,0,0.55) rectangle (1.75,0,0.85);
	\draw[densely dotted, blue,thick] (0.75,0,1.4) --(0.75,0,2) ;
		\filldraw[blue,thick]  (0.75,0,1.4) circle (0.04);
	\end{tikzpicture}.
	\end{equation*}
That is, this action transforms $\Pi_{\sigma}$ as
\begin{equation}\label{braidchargeab}
  B^{[\sigma]}_g(\Pi_{\sigma})=\sum_{h\in G} \tau_{\myinv{g}}(C_{\sigma,h})\otimes \bar{C}_{\sigma,h},
\end{equation}
where $B^{[\sigma]}_g$ stands for braiding with $g$ on one charge $\sigma$ of the pair. If any one of the anyons involved is abelian, i.e. if $\sigma$ is one-dimensional or $g\in Z(G)$, the effect of the braiding is a phase factor: 
$B^{[\sigma]}_g(\Pi_{\sigma})=\tau_{\myinv{g}}(C_{\sigma})\otimes \bar{C}_{\sigma}=(\chi_\sigma(g)/d_\sigma) \Pi_{\sigma}.$

We remark a fundamental point for our work: the relation between braiding and SF. In virtue of Eq.\eqref{braidchargeab} we can state that the factor that relates the action of $\Phi_k\circ \Phi_q$ and $\Phi_{kq}$ over the charge, see Eq.\eqref{eq:symcharge}, is equal to the braiding with the flux $\omega(k,q)\in G$ on that charge. That is, the conjugation by $\omega(k,q)$, $\tau_{\omega(k,q)}$, that defines the braiding in \eqref{braidchargeab} appears in Eq.\eqref{eq:symcharge}.

In general, to detect the effect of the braiding we have to project with the initial state. Using $G$-injectivity, Eq.(\ref{Ginjec}), we observe that
 \begin{align}\label{braid-result}
 \langle   \mathcal{O}^\dagger_{ \sigma} (x,y) \; \mathcal{B}^{[\sigma]}_g \; \mathcal{O}_{ \sigma} (x,y)  \rangle & =  
\begin{tikzpicture}
           \pic at (0.4,0,0.7) {3dpepsshort};
        \pic at (0.4,0.4,0.7) {3dpepsdownshort};
                   \pic at (1.1,0,0.7) {3dpeps};
        \pic at (1.1,0.4,0.7) {3dpepsdown};
   \node[anchor=west] at (-0.4,0.2,0.7)  {$\cdots$};
      \node[anchor=west] at (2.7,0.2,0.7)  {$\cdots$};
         \node[rotate=45] at (1.45,0.4,0.2)  {$\cdots$};
         \node[rotate=45] at (1.45,0,1.2)  {$\cdots$};
           \pic at (1.8,0,0.7) {3dpeps};
        \pic at (1.8,0.4,0.7) {3dpepsdown};
	           \pic at (2.5,0,0.7) {3dpepsshort};
        \pic at (2.5,0.4,0.7) {3dpepsdownshort};
        	\filldraw[draw=blue,fill=white,thick]  (0.55,0,0.7) circle (0.05);
             \filldraw[draw=black,fill=orange,thick] (0.75,0,0.5) rectangle (0.75,0,0.9);
             	\filldraw[blue]  (0.95,0,0.7) circle (0.05);
            \filldraw[draw=black,fill=orange,thick] (0.75,0.4,0.5) rectangle (0.75,0.4,0.9); 
                  \filldraw[draw=black,fill=orange,thick] (2.15,0.4,0.5) rectangle (2.15,0.4,0.9 );  
            \filldraw[draw=black,fill=orange,thick] (2.15,0,0.5) rectangle (2.15,0,0.9);	
  \end{tikzpicture}  \\
& \propto \; \sum_{b\in G} \;
 \begin{tikzpicture}[baseline=-1mm]
 \draw[thick] (-0.4,0.3) rectangle (0.4,-0.3); 
 \draw[->] (-0.4,0.3)--(-0.2,0.3);
  \draw[->] (0.1,0.3)--(0.3,0.3);
    \filldraw[draw=black,fill=orange,thick] (-0.1,-0.4) rectangle (0.1,-0.2);
    \filldraw[draw=black,fill=orange,thick] (-0.1,0.4) rectangle (0.1,0.2);
    \filldraw[draw=blue,fill=white,thick]  (-0.25,-0.3) circle (0.06);
     \node[anchor=north] at (-0.25,-0.3)   {$\myinv{g}$};
     \filldraw[blue]  (0.25,-0.3) circle (0.06);
         \node[anchor=north] at (0.25,-0.4)   {${g}$};
         \node[anchor=west] at (-0.4,0)  {${b}$};
          \filldraw (-0.4,0)  circle (0.04);
 \node[anchor=west] at (0.4,0)  {$\myinv{b}$};
        \filldraw (0.4,0) circle (0.04);
 \end{tikzpicture}
\times
   \begin{tikzpicture}
    \draw[thick] (-0.4,0.3) rectangle (0.4,-0.3); 
    \draw[->] (-0.4,0.3)--(-0.2,0.3);
      \draw[->] (0.4,-0.3)--(0.2,-0.3);
    \filldraw[draw=black,fill=orange,thick] (-0.1,-0.4) rectangle (0.1,-0.2);
    \filldraw[draw=black,fill=orange,thick] (-0.1,0.4) rectangle (0.1,0.2);
      \node[anchor=west] at (-0.4,0)  {${b}$};
          \filldraw[thick] (-0.4,0)  circle (0.04);
 \node[anchor=west] at (0.4,0)  {$\myinv{b}$};
        \filldraw (0.4,0) circle (0.04);
 \end{tikzpicture}, \notag
 \end{align}
where the sum over $b \in G$ is the only part that remains when Eq.(\ref{Gisoconca}) is used to evaluate the overlap. 
It can be shown, whether $G$ is abelian or not, that 
\bed
\langle  \; \mathcal{O}^\dagger_{ \sigma} (x,y) \; \mathcal{B}^{[\sigma]}_g \; \mathcal{O}_{ \sigma} (x,y) \; \rangle / \langle \psi(A)|\psi(A)\rangle = \chi_\sigma(g)/d_\sigma,
\eed
where $\mathcal{B}^{[\sigma]}_g$ stands for the physical action of the braiding. The square of this quantity is the probability of the pair to fuse to the vacuum after braiding, \textit{i.e.} the change in the total charge of the pair. Therefore, braiding allows us to identify the type of a given unknown flux using a probe charge (or a set of them)\cite{Preskill04, Schuch10}. %
We will see in the next subsection how this identification is used to detect the SF class in virtue if its relation with braiding.

\subsection{Identifying the {SF} pattern}

 Given a $G$-isometric {PEPS} with a global on-site symmetry group $Q$, and a permutation action over the anyons $\phi$, we address the problem of resolving between the different possible {SET} phases. For that purpose, we will look for a gauge-invariant quantity which distinguishes between inequivalent cocycles.
An important observation is that $\omega$ itself is not gauge-invariant, as Eq.(\ref{cocylefreedom}) shows; its physical detection would be a meaningless endeavour.  But products of virtual symmetry operators,  
\beq
\lambda(v_{q_1},\dots, v_{q_n}) \equiv v_{q_1}v_{q_2}\cdots v_{q_n} \in G
\eeq
can be gauge invariant if the elements $\{ q_i \in Q \}$ are appropriately chosen. Of course, we already know an important example of gauge-invariant quantity: the result of braiding a flux around a probe charge. We have also mentioned a relation between braiding and symmetry fractionalization. These two facts suggest to construct an operator whose mean value will be analogous to the quantity (\ref{braid-result}) discussed in the previous subsection.

The order parameter we propose is: 
\begin{equation*}\label{orderparameter}
 \Lambda\equiv \langle \;  \mathcal{O}^\dagger_{ \sigma} (x',y)  \; P_{\pi} \; \left( U^{[s_1]}_{q_1}\otimes  \cdots \otimes U^{[s_m]}_{q_m}\right ) \;\mathcal{O}_{ \sigma} (x,y) \; \rangle,
\end{equation*}
where $P_{\pi}$ is an operator representing some permutation $\pi$ of $m$ sites, the upper index $s_j$ denotes the site where $U_{q_j}$ acts, and the indices $x, y$ denote the sites where the charges are created. 
The permutation rearranges the virtual symmetry operators so as to end with a gauge-invariant quantity $\lambda$ acting on the charges analogously as the element $g$ in Eq.\eqref{braid-result}: $\lambda$ plays the role of the flux.
 An interesting identity relates $\Lambda$ with the phase factors resulting from braiding discussed above:
$$\hat{\Lambda} \equiv \Lambda/  \langle \; \mathcal{O}^\dagger_{ \sigma} (x',y) \; P_{\pi} \; \mathcal{O}_{ \sigma} (x,y)\; \rangle = \chi_\sigma(\lambda)/d_\sigma.$$
We remark that the proposed order parameter is \emph{local} in the sense that it can be written as an operator, $X_{\rm local}$, acting on a finite region of the lattice $\Lambda=\langle X_{\rm local} \rangle=\tr[\rho X_{\rm local} ]$. \\

The classification of cocycles and twisting maps is equivalent to the classification of inequivalent group extensions, $E$, of $G$ by $Q$, see Appendix \ref{ap:ext}. This mathematical equivalence has a nice physical interpretation: such extensions correspond to groups associated with the quantum double obtained after gauging the symmetry of the topological phase. Gauging is a transformation to promote a global symmetry into a local one, at the level of Hamiltonians \cite{Jenkins2013} or states \cite{Haegeman14}, ending in a pure topological phase. Since under gauging, inequivalent phases are mapped to inequivalent phases, the classification of {SET} phases is translated to a classification of topological orders; \textit{i.e.} gauged {SET} phases \cite{Barkeshli14}.

Since the class of the cocycle and the twisting map characterise uniquely the extension group $E$ of $Q\cong E/G$ by $G \triangleleft E$, the maps $(\omega, \phi)$ can also be obtained when considering $v_k$ as an element of $E$. The assignment is done by choosing $v_q$ as some element of $E$ for each $q\in Q$ which preserves the quotient map $E\to E/G\cong Q$. This analogy allows us to see $\lambda$ as an invariant quantity of the extension group. Therefore, given ($G,Q, \phi$), the strategy is: $(i)$ we search for all the possible group extensions which characterize all possible symmetry actions, $(ii)$ we look for a gauge-invariant quantity (in the form of $\lambda=\Pi_i v_{q_i}$) that distinguishes between the different extensions, $(iii)$ we design the order parameter $\Lambda$, that is, we identify the appropriate permutation of sites, and apply the suitable symmetry operators, so that the result is equivalent to the braiding of a probe charge with the 'flux' $\lambda \in G$. \\
 \section{Examples}
The SET phases analyzed in this section are introduced in \cite{Hermele14,Tarantino16} as exactly solvable lattice models or in their {PEPS} representation in \cite{Garre17, Williamson17}. %
We include an explicit tensor from  \cite{Garre17} in Appendix \ref{projrep} which corresponds to the first example analyzed in this section. The necessary background on group extensions can be found in Ref.\cite{Adem04}.
%

 \subsection{Toric Code with $\mathbb{Z}_2$ symmetry}\label{TCZ2}

 The Toric Code, $ G = \mathbb{Z}_2 = \{ +1, -1; \times \}$, has three non-trivial anyons: the charge $\sigma$, the flux $m$ and the combination of both. We consider this model with an internal symmetry  $Q = \mathbb{Z}_2 = \{ e, q; q^2 = e \}$ which does not permute the flux and the charge. There are two possible {SF} patterns for the charge as predicted by the identity $H^2(\mathbb{Z}_2,\mathbb{Z}_2)=\mathbb{Z}_2$. The non-trivial {SF} class is characterized by the following projective action on the the charge: 
 $$
(\Phi_q \circ \Phi_q)(C_\sigma)=  -1\times C_\sigma \equiv B^{[\sigma]}_m (C_\sigma).
$$ 
That is, the charge picks up a sign, equivalent to the braiding with the flux, when the symmetry acts twice on it. Notice that because the charge and anticharge are the same particle and they are created together under a local operation, the two minus signs globally cancel out. This is consistent with the fact that globally the symmetry acts linearly. The only non-trivial cocycle here is $\omega(q,q)=-1$. One possible gauge-invariant quantity is $\lambda=\omega(e,e)\omega(q,q)=v^2_{q}$ which gives $-1$ in the non-trivial case and is the identity element for the trivial {SF} class. $\lambda$ is gauge-invariant because $v'^2_{q} = v^2_{q} l^2_q$ and $l_q \in \mathbb{Z}_2$, where of course $l^2_q=e$.  The associated group extensions are $\mathbb{Z}_4$ for the non-trivial case and $\mathbb{Z}_2\times \mathbb{Z}_2$ for the trivial {SF} class. In this example, the gauge-invariant quantity detects whether there are elements of the extension group with order greater than two.
  
We now describe a protocol, and the corresponding diagrams, that distinguishes between these two {SET} phases:
\begin{equation}\label{transprotocol}
   \begin{tikzpicture}[scale=1]
              \pic at (0,0,0.7) {3dpepsshort};
        \pic at (0,1.5,0.7) {3dpepsdownshort};
        \pic at (0.5,0,0) {3dpeps};
        \pic at (0.5,1.5,0) {3dpepsdown};
         \pic at (0.5,0,1.4) {3dpeps};
        \pic at (0.5,1.5,1.4) {3dpepsdown};
                \pic at (1,0,0) {3dpeps};
        \pic at (1,1.5,0) {3dpepsdown};
         \pic at (1,0,1.4) {3dpeps};
           \pic at (1.5,0,0.7) {3dpeps};
        \pic at (1.5,1.5,0.7) {3dpepsdown};
	           \pic at (2,0,0.7) {3dpepsshort};
        \pic at (2,1.5,0.7) {3dpepsdownshort};
          \begin{scope}[canvas is xy plane at z=0]
          \draw[preaction={draw, line width=1.2pt, white},blue,thick] (0.5,1.5,0.7) to  (0.5,1.05,0.7);
          \draw[preaction={draw, line width=1.2pt, white},blue,thick] (1,1.5,0.7) to  (1,1.05,0.7);
          \draw[preaction={draw, line width=1.2pt, white},blue,thick] (1,1.1,0.7) to [out=-90, in=90] (0.5,0,0.7);
          \draw[preaction={draw, line width=1.2pt, white},blue,thick] (0.5,1.1,0.7) to [out=-90, in=90] (1,0,0.7);
        \end{scope}   
        \draw[thick] (1,1.5,1.4)--(1,1.31,1.4);
    \begin{scope}[canvas is zx plane at y=1.5]
      \draw[preaction={draw, line width=1.2pt, white},thick] (0.9,1)--(1.9,1);
      \draw[preaction={draw, line width=1.2pt, white},thick] (1.4,0.6)--(1.4,1.4);
      \filldraw (1.4,1) circle (0.07);
    \end{scope} 
         \pic at (1,0,0.7) {3dpepsp};
        \pic at (1,1.5,0.7) {3dpepsdownp};
        \pic at (0.5,0,0.7) {3dpepsp};
        \pic at (0.5,1.5,0.7) {3dpepsdownp};
             \filldraw[draw=black,fill=orange,thick] (1.25,0,0.5) rectangle (1.25,0,0.9 );
            \filldraw[draw=black,fill=orange,thick] (0.25,1.5,0.5) rectangle (0.25,1.5,0.9); 
              \filldraw[draw=black,fill=orange,thick] (1.75,1.5,0.5) rectangle (1.75,1.5,0.9 );  
            \filldraw[draw=black,fill=orange,thick] (1.75,0,0.5) rectangle (1.75,0,0.9  );

              \filldraw[draw=black,fill=purple,thick] (0.52,0.95,0.7) circle (0.07);
              \filldraw[draw=black,fill=purple,thick] (0.98,0.95,0.7) circle (0.07);
               \node at (4,1.6,0) {{\sf (i)  Create the excited state with}};
               \node at (4.1,1.3,0) {{\sf  2 charges: $\mathcal{O}_{\sigma} (x,y) |{\rm TC} \rangle$}};
               \node at (4,0.9,0) {{\sf (ii) Apply the on-site symmetry}};
                 \node at (4,0.6,0) {{\sf operators: $U^{\otimes 2}_q$}};
                 \node at (4,0.2,0) {{\sf (iii) Permute sites: $P_{(12)}$}};
                   \node at (4,-0.2,0) {{\sf (iv) Project onto $\langle {\rm TC}|  \mathcal{O}^\dagger_{\sigma} (x',y)$}};
        \end{tikzpicture},
        \end{equation}
where $P_{(12)}$ denotes the permutation of the two sites where the symmetry acts. 
We are assuming the contractions of the physical indices, between the bra and the ket layers,  for the sake of clarity we will not draw them.
The order parameter associated to this protocol is the following:
 $$
 \Lambda = \langle {\rm TC}| \mathcal{O}^\dagger_{\sigma} (x',y) P_{(12)} U^{\otimes 2}_q  \mathcal{O}_{\sigma} (x,y)|{\rm TC}\rangle.
 $$
Notice that the charges appearing in the ket and the bra are not placed on the same links: $x'\neq x$. 
We compute $\hat{\Lambda}$ using the advantages of calculating local expectation values in the $G$-isometric PEPS picture (see explanation below Eq.\eqref{Gisoconca}), we obtain the following:
\begin{equation*}
\hat{\Lambda}=
\frac{1}{|G|}
\sum_{b\in G}\;
   \begin{tikzpicture}[scale=1]
                \node[anchor= south west] at (0.5,0.75,-0.2) {$\myinv{b}$};
        \filldraw (0.5,0.75,-0.2) circle (0.05);
         \draw[->] (0.5,1,0.5)--(0.5,1,0.2);
                              \node[anchor=south west] at (1,0.75,-0.2) {$\myinv{b}$};
        \filldraw (1,0.75,-0.2) circle (0.05);
        \draw[->] (1,1,0.5)--(1,1,0.2);
\draw[thick] (0.5,1,0.7)--(0.5,1,-0.2)--(0.5,0,-0.2)--(0.5,0,0.7);
\draw[thick] (1,1,0.7)--(1,1,-0.2)--(1,0,-0.2)--(1,0,0.7);
\draw[preaction={draw, line width=1.2pt, white},thick] (0,0,0.7) rectangle (1.5,1,0.7);
          \draw[preaction={draw, line width=1.2pt, white},thick, blue] (1,1,0.7) to [out=-90, in=90] (0.5,0,0.7);
          \draw[preaction={draw, line width=1.2pt, white},thick, blue] (0.5,1,0.7) to [out=-90, in=90] (1,0,0.7);  
              \filldraw[draw=black,fill=purple,thick] (0.55,0.75,0.7) circle (0.07);
              \filldraw[draw=black,fill=purple,thick] (0.95,0.75,0.7) circle (0.07);
              \draw[preaction={draw, line width=1.2pt, white},thick] (0.5,1,0.7)--(0.5,1,1.6)--(0.5,0,1.6)--(0.5,0,0.7);
              \draw[preaction={draw, line width=1.2pt, white},thick] (1,1,0.7)--(1,1,1.6)--(1,0,1.6)--(1,0,0.7);
         \pic at (1,0,0.7) {3dpepsp};
        \pic at (1,1,0.7) {3dpepsdownp};
        \pic at (0.5,0,0.7) {3dpepsp};
        \pic at (0.5,1,0.7) {3dpepsdownp};
                             \filldraw[draw=black,fill=orange,thick] (1.25,0,0.5) rectangle (1.25,0,0.9 );
            \filldraw[draw=black,fill=orange,thick] (0.25,1,0.5) rectangle (0.25,1,0.9);
                        \node[anchor= north east] at (0.5,0.25,1.6) {${b}$};
        \filldraw (0.5,0.25,1.6) circle (0.05);
         \draw[->] (0.5,0,0.7)--(0.5,0,1.3);
                                \node[anchor= north east] at (1,0.25,1.6) {${b}$};
        \filldraw (1,0.25,1.6) circle (0.05); 
         \draw[->] (1,0,0.7)--(1,0,1.3);
                                  \node[anchor=west] at (1.5,0.7,0.7) {$\myinv{b}$};
        \filldraw (1.5,0.7,0.7) circle (0.05);
         \draw[->] (1.5,0.7,0.7) --(1.5,0.3,0.7);
 \node[anchor=east] at (0,0.3,0.7) {${b}$};
        \filldraw (0,0.3,0.7) circle (0.05);
         \draw[->] (0,0.5,0.7)--(0,0.7,0.7);
\end{tikzpicture}
\times
\begin{tikzpicture}[scale=1]    
 \draw[preaction={draw, line width=1.2pt, white},thick] (1.8,0,0.7) rectangle (2.6,1,0.7);
 \draw[->] (1.8,1,0.7)--(2,1,0.7);
 \draw[->] (2.6,0,0.7)--(2.4,0,0.7);
              \filldraw[draw=black,fill=orange,thick] (2.2,1,0.5) rectangle (2.2,1,0.9 );  
            \filldraw[draw=black,fill=orange,thick] (2.2,0,0.5) rectangle (2.2,0,0.9  );
             \node[anchor=east] at (1.8,0.5,0.7) {${b}$};
        \filldraw (1.8,0.5,0.7) circle (0.05);
         \node[anchor=west] at (2.6,0.5,0.7) {$\myinv{b}$};
        \filldraw (2.6,0.5,0.7) circle (0.05);
   \end{tikzpicture},
  \end{equation*}
where the sum over $b\in \mathbb{Z}_2$ comes from the concatenation of the non-permuted sites. 
We notice that the position of the two charges placed on the rightmost edges, one on top of the other, in \eqref{transprotocol} can be changed without altering the value of $\hat{\Lambda}$.  
Using Eq.(\ref{Ginjec}) and Eq.\eqref{locsym} we get:
\begin{align}\label{loopcalc}
\hat{\Lambda} = \frac{1}{|G|^3}&\sum_{a,b,c\in G} 
  \begin{tikzpicture}[baseline=+5mm]
        \draw[thick, blue] (1.5,1) to [out=-90, in=90] (0.6,0);
        \draw[thick,blue] (1.4,1) to [out=-90, in=90] (0.5,0);
       \draw[preaction={draw, line width=1.2pt, white},thick,blue] (0.6,1) to [out=-90, in=90] (1.5,0);
        \draw[preaction={draw, line width=1.2pt, white},thick,blue] (0.5,1) to [out=-90, in=90] (1.4,0);
        \draw[thick] (0.6,0)--(1.4,0);
        \draw[thick] (1.5,0)--(1.9,0);
        \draw[thick] (1.9,1)--(1.9,0);
        \draw[thick] (0.6,1)--(1.4,1);
        \draw[thick] (1.5,1)--(1.9,1);
        \draw[thick] (0.5,1)--(0,1);
        \draw[thick] (0.5,0)--(0,0);
        \draw[thick] (0,1)--(0,0);
        \node[anchor=south] at (0.35,1) {$\myinv{v_{q}}$};
        \filldraw[draw=black,fill=red,thick] (0.4,1) circle (0.05);
         \node[anchor=east] at (0.7,0.7) {$\myinv{a}$};
        \filldraw (0.58,0.75) circle (0.05);
        \node[anchor=west] at (1.4,0.7) {$c$};
        \filldraw (1.43,0.75) circle (0.05);
        \node[anchor=east] at (0.7,0.2) {$\myinv{c}$};
        \filldraw (0.58,0.25) circle (0.05);
        \node[anchor=west] at (1.4,0.2) {$a$};
        \filldraw (1.43,0.25) circle (0.05);
        \node[anchor=south] at (0.8,1) {$v_{q}$};
        \filldraw[draw=black,fill=red,thick] (0.8,1) circle (0.05);
         \draw[->] (1.2,0)--(1.0,0);
         \draw[->] (0.9,1)--(1.05,1);
         \node[anchor=south] at (1.3,1) {$\myinv{v_{q}}$};
        \filldraw[draw=black,fill=red,thick] (1.3,1) circle (0.05);
        \node[anchor=south,thick] at (1.9,1) {$v_{q} $};
        \filldraw[draw=black,fill=red] (1.7,1) circle (0.05);
        \node[anchor=east] at (0,0.5) {$b$};
        \draw[->] (0,0)--(0,0.3);
        \filldraw (0,0.5) circle (0.05);
        \node[anchor=west,thick] at (1.9,0.5) {$\myinv{b}$};
        \filldraw (1.9,0.5) circle (0.05);
        \draw[->] (1.9,0.9)--(1.9,0.7);
        \filldraw[draw=black,fill=orange,thick] (1.65,-0.1) rectangle (1.85,0.1);
        \filldraw[draw=black,fill=orange,thick] (0.05,0.9) rectangle (0.25,1.1);
         \end{tikzpicture}  \times  \;
  \begin{tikzpicture}
       \draw[thick] (0,0) rectangle (1,1);
       \draw[->] (0,1)--(0.25,1);
       \draw[->] (1,0)--(0.75,0);
        \node[anchor=west] at (0,0.5) {$b$};
        \filldraw (0,0.5) circle (0.05);
       \node[anchor=east] at (1.1,0.5) {$\myinv{b}$};
        \filldraw (1,0.5) circle (0.05);
        \filldraw[draw=black,fill=orange,thick] (0.4,0.9) rectangle (0.6,1.1);
      \filldraw[draw=black,fill=orange,thick] (0.4,-0.1) rectangle (0.6,0.1);
     \end{tikzpicture} \notag \\
 &  \times \;
    \begin{tikzpicture}
          \draw[preaction={draw, line width=1.2pt, white},thick,blue] (0,0,-1) to [out=90, in=-90] (1,1,-1);
          \draw[preaction={draw, line width=1.2pt, white},thick, blue] (1,0,-1) to [out=90, in=-90] (0,1,-1);
           \draw[thick] (0,0,0)--(0,1,0);
       \draw[preaction={draw, line width=1.2pt, white},thick] (1,0,0)--(1,1,0);
         \node[anchor=west] at (0.05,0.8,-1) {$\myinv{a}$};
        \filldraw (0.05,0.8,-1) circle (0.05);
        \node[anchor=east] at (0.2,0.3,-1) {$\myinv{c}$};
        \filldraw (0.05,0.2,-1) circle (0.05);
        \node[anchor=east] at (0,0.6,0) {$b$};
        \filldraw (0,0.6,0) circle (0.05);
        \draw[->] (0,0,0)--(0,0.4,0);
        \node[anchor=east] at (1,0.6,0) {$b$};
        \filldraw (1,0.6,0) circle (0.05);
        \draw[->] (1,0,0)--(1,0.4,0);
   \begin{scope}[canvas is zx plane at y=0]
     \draw[thick] (0,0)--(-1,0);
     \draw[thick] (0,1)--(-1,1);
   \end{scope} 
  \begin{scope}[canvas is zx plane at y=1]
      \draw[thick] (0,0)--(-1,0);
      \draw[thick] (0,1)--(-1,1);
   \end{scope} 
            \node[anchor=south] at (0,1,-0.5) {$v_{q}$};
        \filldraw[draw=black,fill=red,thick] (0,1,-0.5) circle (0.06);
      \node[anchor=south] at (1,1,-0.5) {$v_{q}$};
        \filldraw[draw=black,fill=red,thick] (1,1,-0.5) circle (0.06);
      \end{tikzpicture}
      \; \times  
        \begin{tikzpicture}
        \draw[thick] (0,0,0)--(0,1,0);
         \draw[thick] (1,0,0)--(1,1,0);
          \draw[preaction={draw, line width=1.2pt, white},thick,blue] (0,0,1) to [out=90, in=-90] (1,1,1);
          \draw[preaction={draw, line width=1.2pt, white},thick,blue] (1,0,1) to [out=90, in=-90] (0,1,1);
         \node[anchor=east] at (0.05,0.8,1) {$a$};
        \filldraw (0.05,0.8,1) circle (0.05);
        \node[anchor=east] at (0.05,0.2,1) {$c$};
        \filldraw (0.05,0.2,1) circle (0.05);
        \node[anchor=west] at (0,0.4,0) {$\myinv{b}$};
        \filldraw (0,0.4,0) circle (0.05);
        \draw[->] (0,0.8,0)--(0,0.6,0);
        \node[anchor=east] at (1.1,0.5,0) {$\myinv{b}$};
        \filldraw (1,0.5,0) circle (0.05);
        \draw[->] (1,0.5,0)--(1,0.2,0);
     \begin{scope}[canvas is zx plane at y=0]
     \draw[thick] (0,0)--(1,0);
     \draw[thick] (0,1)--(1,1);
      \end{scope} 
     \begin{scope}[canvas is zx plane at y=1]
      \draw[thick] (0,0)--(1,0);
      \draw[thick] (0,1)--(1,1);
     \end{scope} 
     \node[anchor=south] at (0,1,0.5) {$\myinv{v_{q}}$};
        \filldraw[draw=black,fill=red,thick] (0,1,0.5) circle (0.06);
      \node[anchor=south] at (1,1,0.5) {$\myinv{v_{q}}$};
        \filldraw[draw=black,fill=red,thick] (1,1,0.5) circle (0.06);
   \end{tikzpicture}.
 \end{align}
 The first factor in the sum is $\tr[  C^{[u]}_\sigma v^{-1}_q a^{-1}  c v_qb^{-1} C^{[d]}_\sigma a v_q v^{-1}_qc^{-1}  b ]$ and simplifies to  $\tr[  C^{[u]}_\sigma {b}^{-1} (a {c}^{-1})^{-1}  C^{[d]}_\sigma (a {c}^{-1}) b ]$ using that $v^{-1}_q (a^{-1}  c) v_q=a^{-1}  c$ (because $\phi_q=\id$ in this case) and that $a,b,c$ commute pairwise since they belong to $\mathbb{Z}_2$. We see that all symmetry operators cancel out. This simplification allows us to diagrammatically express that factor as one of the loops that we will relate afterwards with Eq.\eqref{braid-result}:
  \begin{equation*}
 \begin{tikzpicture}
         \draw[thick, blue] (1.5,1) to [out=-90, in=90] (0.6,0);
        \draw[thick,blue] (1.4,1) to [out=-90, in=90] (0.5,0);
       \draw[preaction={draw, line width=1.2pt, white},thick,blue] (0.6,1) to [out=-90, in=90] (1.5,0);
        \draw[preaction={draw, line width=1.2pt, white},thick,blue] (0.5,1) to [out=-90, in=90] (1.4,0);
        \draw[thick] (0.6,0)--(1.4,0);
        \draw[thick] (1.5,0)--(1.9,0);
        \draw[thick] (1.9,1)--(1.9,0);
        \draw[thick] (0.6,1)--(1.4,1);
        \draw[thick] (1.5,1)--(1.9,1);
        \draw[thick] (0.5,1)--(0,1);
        \draw[thick] (0.5,0)--(0,0);
        \draw[thick] (0,1)--(0,0);
        \node[anchor=south] at (0.35,1) {$\myinv{v_{q}}$};
        \filldraw[draw=black,fill=red,thick] (0.4,1) circle (0.05);
         \node[anchor=east] at (0.7,0.7) {$\myinv{a}$};
        \filldraw (0.58,0.75) circle (0.05);
        \node[anchor=west] at (1.4,0.7) {$c$};
        \filldraw (1.43,0.75) circle (0.05);
        \node[anchor=east] at (0.7,0.2) {$\myinv{c}$};
        \filldraw (0.58,0.25) circle (0.05);
        \node[anchor=west] at (1.4,0.2) {$a$};
        \filldraw (1.43,0.25) circle (0.05);
        \node[anchor=south] at (0.8,1) {$v_{q}$};
        \filldraw[draw=black,fill=red,thick] (0.8,1) circle (0.05);
         \draw[->] (1.2,0)--(1.0,0);
         \draw[->] (0.9,1)--(1.05,1);
         \node[anchor=south] at (1.3,1) {$\myinv{v_{q}}$};
        \filldraw[draw=black,fill=red,thick] (1.3,1) circle (0.05);
        \node[anchor=south,thick] at (1.9,1) {$v_{q} $};
        \filldraw[draw=black,fill=red] (1.7,1) circle (0.05);
        \node[anchor=east] at (0,0.5) {$b$};
        \draw[->] (0,0)--(0,0.3);
        \filldraw (0,0.5) circle (0.05);
        \node[anchor=west,thick] at (1.9,0.5) {$\myinv{b}$};
        \filldraw (1.9,0.5) circle (0.05);
        \draw[->] (1.9,0.9)--(1.9,0.7);
        \filldraw[draw=black,fill=orange,thick] (1.65,-0.1) rectangle (1.85,0.1);
        \filldraw[draw=black,fill=orange,thick] (0.05,0.9) rectangle (0.25,1.1);
         \end{tikzpicture} =
             \begin{tikzpicture}[baseline=-2mm]
 \draw[thick] (-0.4,0.3) rectangle (0.4,-0.3);
 \draw[->] (-0.4,0.3)--(-0.2,0.3); 
  \draw[->] (0.1,0.3)--(0.3,0.3); 
    \filldraw[thick, draw=black,fill=orange] (-0.1,-0.4) rectangle (0.1,-0.2);
    \filldraw[thick, draw=black,fill=orange] (-0.1,0.4) rectangle (0.1,0.2);
    \filldraw[draw=blue,fill=white,thick]  (-0.25,-0.3) circle (0.06);
   \node[anchor=north ] at (-0.5,-0.3)  {$\myinv{(a\myinv{c})}$}; 
     \filldraw[blue,thick]  (0.25,-0.3) circle (0.06);
     \node[anchor=north] at (0.5,-0.3)  {${a\myinv{c}}$}; 
         \node[anchor=west] at (-0.4,0)  {${b}$};
          \filldraw (-0.4,0)  circle (0.05);
 \node[anchor=west] at (0.4,0)  {$\myinv{b}$};
        \filldraw (0.4,0) circle (0.05);
 \end{tikzpicture}.
 \end{equation*}
The second factor of Eq.\eqref{loopcalc} is $\tr[  \bar{C}^{[u]}_\sigma  b^{-1} \bar{C}^{[d]}_\sigma b ]$. Importantly, the product of the first and the second factors is equal to the result of the braiding detection of a charge $\sigma$ and a flux $g$, see Eq.(\ref{braid-result}), with the identification of $g\equiv a {c}^{-1}$. 

\

The dependence on the {SF} class lies in the remaining two loops, both equal to $\tr[a v^{-1}_q b^{-1} c v^{-1}_q b^{-1}]= \tr[a c v^{-2}_q]= |G|\delta_{ac,v^2_q}$. 
These factors can only be non-zero if $ac^{-1}= v^2_q \equiv \lambda$ since $c=c^{-1}$ because $c\in \mathbb{Z}_2$. 
Notice that the two loops implying the previous identity are not part of the plane formed by the charges and the permutation in Eq.(\ref{transprotocol}): it is an intrinsic 2D effect of $G$-injective PEPS. In summary, we can write
\begin{align*}
\hat{\Lambda}=& \;\frac{1}{2} \sum_{b\in \mathbb{Z}_2} \tr[ C^{[u]}_\sigma  v^{-2}_q b^{-1} C^{[d]}_\sigma b v^2_q  ]\times \tr[  \bar{C}^{[u]}_\sigma  b^{-1} \bar{C}^{[d]}_\sigma b ]  \\
=&\; \frac{1}{2} \sum_{b\in \mathbb{Z}_2} \;
 \begin{tikzpicture}[baseline=-1mm]
 \draw[thick] (-0.4,0.3) rectangle (0.4,-0.3); 
  \draw[->] (-0.4,0.3)--(-0.2,0.3); 
  \draw[->] (0.1,0.3)--(0.3,0.3); 
    \filldraw[draw=black,fill=orange,thick] (-0.1,-0.4) rectangle (0.1,-0.2);
    \filldraw[draw=black,fill=orange,thick] (-0.1,0.4) rectangle (0.1,0.2);
    \filldraw[draw=blue,fill=white,thick]  (-0.25,-0.3) circle (0.06);
     \filldraw[blue]  (0.25,-0.3) circle (0.06);
        \node[anchor=north ] at (-0.4,-0.3)  {$\myinv{\lambda}$}; 
     \node[anchor=north] at (0.4,-0.3)  {${\lambda}$}; 
         \node[anchor=west] at (-0.4,0)  {${b}$};
          \filldraw (-0.4,0)  circle (0.05);
 \node[anchor=west] at (0.4,0)  {$\myinv{b}$};
        \filldraw (0.4,0) circle (0.05);
 \end{tikzpicture}
\times
   \begin{tikzpicture}
    \draw[thick] (-0.4,0.3) rectangle (0.4,-0.3); 
     \draw[->] (-0.4,0.3)--(-0.2,0.3); 
  \draw[->] (0.4,-0.3)--(0.2,-0.3); 
    \filldraw[draw=black,fill=orange,thick] (-0.1,-0.4) rectangle (0.1,-0.2);
    \filldraw[draw=black,fill=orange,thick] (-0.1,0.4) rectangle (0.1,0.2);
      \node[anchor=west] at (-0.4,0)  {${b}$};
          \filldraw (-0.4,0)  circle (0.05);
 \node[anchor=west] at (0.4,0)  {$\myinv{b}$};
        \filldraw (0.4,0) circle (0.05);
 \end{tikzpicture}
=\chi_{\sigma}(\lambda). 
\end{align*}
The value of $\hat{\Lambda}$ is equal to $+ 1$ or $-1$, depending on the {SF} class; trivial or non-trivial respectively.      
\\

Let us compare our approach with a discrimination based on mapping a 2D {SET} system to a cylinder, and using {SPT} classification tools. Concretely, for each anyon type one constructs a 1D system, and places an anyon and its antiparticle on each edge of the cylinder. Then one could wonder if the symmetric projective action on the bonds in those chains, characterized by SPT phases in 1D, is equivalent to the SF pattern of the corresponding anyons. The algebraic object that classifies SPT phases in 1D is $H^2(Q,U(1))$, whereas in SF it is $H^2(Q,G)$: the topological order constraints the values of the projective actions from $U(1)$ to $G$ when $G$ is abelian. Therefore for each anyon of the TC with a $\mathbb{Z}_2$ symmetry, the corresponding SPT phase is in $H^2(\mathbb{Z}_2,U(1))$, which is trivial. Therefore \emph{no} signature of the SF pattern of the charge can be found with compactification. In general for symmetries coming from a cyclic group, the 1D {SPT} phase is always trivial: $H^2(\mathbb{Z}_n,U(1))=1$, so any non-trivial SF pattern of a cyclic group have no 1D SPT analogue (this situation includes examples C,D and E below). In the next example we show that even when there are non-trivial {SPT} phases after compactification, they do not fully resolve between all the SF patterns of the anyons.

\subsection{ Toric Code with $\mathbb{Z}_2 \times \mathbb{Z}_2$ symmetry }

\begin{figure}[ht!]
\begin{center}
\includegraphics[scale=1]{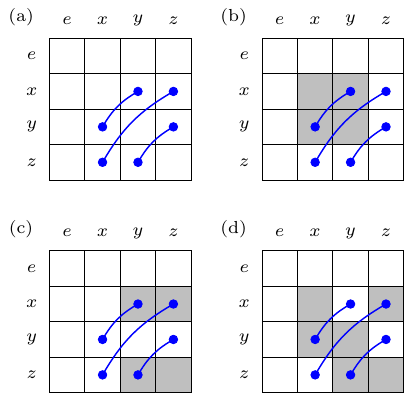}
\caption{(Color online) Matrix representation, $(M)_{q,k}=\omega(q,k)$, of the cocycles related to the four $\mathbb{Z}_2\times \mathbb{Z}_2$ {SF} patterns in the TC. The grey shaded areas correspond to the cocycle value $-1$ and the white to $+1$ (the matrices are shown for a specific gauge choice). (a) corresponds to $\mathbb{Z}^3_2$, (b) to $\mathbb{Z}_4 \times \mathbb{Z}_2$ (c) to $D_8$ and (d) to $Q_8$.  The order parameter (\ref{eq:z2z2z2}), the set made of the lower 3 diagonal elements, is distinct for each phase. In turn, the {SPT}-induced order parameter, the set of product of upper and lower diagonal elements identifies (a) with (b) and (c) with (d), blue line connecting $\omega(k,q)$ and $\omega^{-1} (q,k)$ (the concrete values are given in Table \ref{tablev}). }
\label{fig:cocycleconstruction}
\end{center}
\end{figure}

The symmetry $\mathbb{Z}_2 \times \mathbb{Z}_2=\{e,x,y,z\} \subset SO(3)$, considered as $\pi$ rotations over each axis, acting on the Toric Code gives four inequivalent patterns of {SF} on the charge (when the {SF} of the flux is trivial and there is no permutation of anyons). There are three non-trivial {SF} classes, associated with the group extensions $\mathbb{Z}_4\times \mathbb{Z}_2, D_8,Q_8$ and one trivial class, associated with $\mathbb{Z}_2\times\mathbb{Z}_2\times\mathbb{Z}_2\equiv \mathbb{Z}^3_2$. These four classes come from $H^2( Q, G ) = H^2(\mathbb{Z}_2 \times \mathbb{Z}_2 ,\mathbb{Z}_2)=\mathbb{Z}^3_2$ when one has incoporated the redundancy of cocycles by relabeling the elements of $\mathbb{Z}_2 \times \mathbb{Z}_2$ (see Apendix \ref{CohomoTCZ2Z2}). 
The gauge-invariant quantity that identifies any of the four {SF} classes is $v^2_q$ for each $q=x,y,z$. We are led to choose our order parameter to be the triple
\begin{equation}\label{eq:z2z2z2} 
\lambda = \{v^2_x,v^2_z,v^2_y\}
\end{equation} 
(see Table \ref{tablev} and Fig. \ref{fig:cocycleconstruction}). The scheme presented in Section \ref{TCZ2} can be used in order to calculate each element of the triple (the three non-trivial elements of the symmetry group).

 \begin{table}[htb]
\centering
\begin{tabular}{c|c|c|c|c|}
 & $\mathbb{Z}^3_2$ & $\mathbb{Z}_4\times \mathbb{Z}_2$ & $D_8$ & $Q_8$ \\ \hline 
$\{v^2_x,v^2_z,v^2_y\}$ & $\{1,1,1\}$ & $\{1,-1,-1\}$ & $\{1,1,-1\}$ & $\{-1,-1,-1\}$ \\  \hline
$v_xv_zv^{-1}_x v^{-1}_z$ & $+1$ & $+1$ & $-1$ & $-1$ \\ \hline
\end{tabular}
\caption{Comparison between the values of the gauge-invariant quantities.}
\label{tablev}
 \end{table}
 
Let us compare this order parameter with the analogous quantity used in the detection of {1D} systems invariant under the symmetry  $\mathbb{Z}_2 \times \mathbb{Z}_2 \subset SO(3)$. 
In that case the different SPT phases are classified by $H^2(\mathbb{Z}_2 \times \mathbb{Z}_2,U(1))=\mathbb{Z}_2$,
 predicting two non-equivalent phases where the non-trivial one (the trivial corresponds to a product state) is the Haldane phase \cite{Pollmann12}. 

The order parameter for the effective 1D system obtained when putting the system around a ( long ) cylinder is \cite{Haegeman12, Pollmann12}:
$$
v_xv_zv^{-1}_x v^{-1}_z=\omega(x,z)v_{xz}v^{-1}_{zx}\omega^{-1}(z,x)=\omega(x,z)\omega^{-1}(z,x).
$$
A comparison between Table \ref{tablev} and Fig. \ref{fig:cocycleconstruction} allows to contrast the {SPT} approach with ours: the former doesn't resolve between the 4 {SF} phases, whereas the latter does. The {SPT} approach is only able to discriminate between the sets corresponding to $\{ \mathbb{Z}^3_2, \mathbb{Z}_4\times \mathbb{Z}_2\}$ and $\{D_8,Q_8\}$.
        
\subsection{$\mathcal{D}$($\mathbb{Z}_p$) with $\mathbb{Z}_p$ symmetry, $p$ prime}\label{secq}

We are going to work with the following presentation of the cyclic group with $p$ elements: $\mathbb{Z}_p = \{ \langle q \rangle: q^p = e \}$ 
Since $p$ is prime, the homomorphism from $Q=\mathbb{Z}_p$ to ${\rm Aut}(G) = {\rm Aut}(\mathbb{Z}_p)\cong \mathbb{Z}_{p-1}$ is trivial ($\phi_q=\id$\footnote{
Since $\phi$ is an homomorphism from $\mathbb{Z}_p \to \mathbb{Z}_{p-1}$ it satisfies $\id=\phi^{p-1}_q=\phi_{q^{p-1}}$ for any $q\in \mathbb{Z}_p$ but since $p-1$ and $p$ are coprimes $\{ q^{p-1}\}_{q\in \mathbb{Z}_p}= \mathbb{Z}_p$ so $\phi_q=\id , \; \forall q\in \mathbb{Z}_p$. 
}) 
so there is no permutation of anyons. This implies that the possible phases are distinguished only by the inequivalent {SF} patterns on the charges.
The cohomological classification gives $H^2(\mathbb{Z}_p,\mathbb{Z}_p)\cong \mathbb{Z}_p$ but there are only two group extensions $\mathbb{Z}_p \times \mathbb{Z}_p$ and $\mathbb{Z}_{p^2}$. There are $p-1$ non-trivial inequivalent cocycles corresponding to $\mathbb{Z}_{p^2}$ that can be related by relabelling the symmetry operators. That is, $\omega \not \equiv \omega'$ but $\omega' \equiv \omega \circ (\rho\times \rho)$ where $\rho \in {\rm Aut}(Q)$. The non-trivial {SF} class is characterized by a projective action on all non-trivial charges of the model.
The quantity we want to measure is 
$$\lambda=v^p_q=\omega(q,e)\omega(q,q)\omega(q,q^2)\cdots \omega(q,q^{p-1}), $$ 
which is gauge-invariant because of Eq.(\ref{cocylefreedom}) since  $v'^p_q=v^p_q l^p_q=v^p_q$ whenever $l_q\in \mathbb{Z}_p$. The values are $v^p_q=e$ if it is the trivial cocycle (group extension $\mathbb{Z}_p \times \mathbb{Z}_p$) or $v^p_q =\alpha$
where $\alpha\in [1,\dots,p-1]$ represents the $p-1$ inequivalent non-trivial cocycles. $\lambda$ measures if the group extension has elements of order greater than $p$.
To construct the order parameter we apply $U^{\otimes p}_q$ and the permutation $(12\cdots p)$ to $p$ consecutive sites on the state with a pair of charges and then we project onto the state with the pair of charges placed as in the previous example:$$\Lambda \equiv  \langle \;  \mathcal{O}^\dagger_{\sigma}(x',y) \;P_{(1\cdots p)}\; U^{\otimes p}_q \; \mathcal{O}_{\sigma} (x,y) \; \rangle .$$ For instance, in the case $p=3$ the order parameter would correspond to
   \begin{equation*} 
   \Lambda=\begin{tikzpicture}[scale=1]
        \pic at (1,0,0) {3dpeps};
        \pic at (1,1,0) {3dpepsdown}; 
        \pic at (0.5,0,0) {3dpeps};
        \pic at (0.5,1,0) {3dpepsdown};
         \pic at (1.5,0,0) {3dpeps};
        \pic at (1.5,1,0) {3dpepsdown};
                   \pic at (0,0,0.7) {3dpepsshort};
        \pic at (0,1,0.7) {3dpepsdownshort};
            \pic at (2,0,0.7) {3dpeps};
       \pic at (2,1,0.7) {3dpepsdown};
          \begin{scope}[canvas is xy plane at z=0]
          \draw[preaction={draw, line width=1.2pt, white},thick,blue] (1.5,1,0.7) to [out=-90, in=90] (0.5,0,0.7);
          \draw[preaction={draw, line width=1.2pt, white},thick,blue] (0.5,1,0.7) to [out=-90, in=90] (1,0,0.7);
          \draw[preaction={draw, line width=1.2pt, white},thick,blue] (1,1,0.7) to [out=-90, in=90] (1.5,0,0.7);
        \end{scope}
         \filldraw[draw=black,fill=purple,thick] (1.47,0.85,0.7) circle (0.06);
         \filldraw[draw=black,fill=purple,thick] (1.03,0.85,0.7) circle (0.06);
         \filldraw[draw=black,fill=purple,thick] (0.53,0.85,0.7) circle (0.06);
            \pic at (1.5,0,0.7) {3dpepsp};
        \pic at (1.5,1,0.7) {3dpepsdownp};
                \pic at (1,0,0.7) {3dpepsp};
        \pic at (1,1,0.7) {3dpepsdownp}; 
                \pic at (0.5,0,0.7) {3dpepsp};
        \pic at (0.5,1,0.7) {3dpepsdownp};
         \pic at (0.5,0,1.4) {3dpeps};
        \pic at (0.5,1,1.4) {3dpepsdown};
                \pic at (1,0,1.4) {3dpeps};
                \pic at (1.5,0,1.4) {3dpeps};
        \draw[thick] (1,1,1.4)--(1,0.81,1.4);
    \begin{scope}[canvas is zx plane at y=1]
      \draw[preaction={draw, line width=1.2pt, white},thick] (0.9,1)--(1.9,1);
      \draw[preaction={draw, line width=1.2pt, white},thick] (1.4,0.6)--(1.4,1.4);
      \filldraw (1.4,1) circle (0.07);
    \end{scope} 
            \draw[thick] (1.5,1,1.4)--(1.5,0.81,1.4);
    \begin{scope}[canvas is zx plane at y=1]
      \draw[preaction={draw, line width=1.2pt, white},thick] (0.9,1.5)--(1.9,1.5);
      \draw[preaction={draw, line width=1.2pt, white},thick] (1.4,1.1)--(1.4,1.9);
      \filldraw (1.4,1.5) circle (0.07);
    \end{scope} 
             \filldraw[draw=black,fill=orange,thick] (1.75,0,0.5) rectangle (1.75,0,0.9);
            \filldraw[draw=black,fill=orange,thick] (0.25,1,0.5) rectangle (0.25,1,0.9); 
          \pic at (2.5,0,0.7) {3dpepsshort};
        \pic at (2.5,1,0.7) {3dpepsdownshort};
             \filldraw[draw=black,fill=orange,thick] (2.25,0,0.5) rectangle (2.25,0,0.9);
            \filldraw[draw=black,fill=orange,thick] (2.25,1,0.5) rectangle (2.25,1,0.9); 
\end{tikzpicture}.
\end{equation*}
Using Eq.(\ref{Gisoconca}):
\begin{equation*}
\hat{\Lambda}=
\frac{1}{|G|}
\sum_{b\in G}\;
   \begin{tikzpicture}[scale=1]
                \node[anchor= south west] at (0.5,0.75,-0.2) {$\myinv{b}$};
        \filldraw (0.5,0.75,-0.2) circle (0.05);
         \draw[->] (0.5,1,0.5)--(0.5,1,0.2);
                      \node[anchor=south west] at (1,0.75,-0.2) {$\myinv{b}$};
        \filldraw (1,0.75,-0.2) circle (0.05);
          \draw[->] (1,1,0.5)--(1,1,0.2);
                              \node[anchor=south west] at (1.5,0.75,-0.2) {$\myinv{b}$};
        \filldraw (1.5,0.75,-0.2) circle (0.05);
          \draw[->] (1.5,1,0.5)--(1.5,1,0.2);
\draw[thick] (0.5,1,0.7)--(0.5,1,-0.2)--(0.5,0,-0.2)--(0.5,0,0.7);
\draw[thick] (1,1,0.7)--(1,1,-0.2)--(1,0,-0.2)--(1,0,0.7);
\draw[thick] (1.5,1,0.7)--(1.5,1,-0.2)--(1.5,0,-0.2)--(1.5,0,0.7);
\draw[preaction={draw, line width=1.2pt, white},thick] (0,0,0.7) rectangle (2,1,0.7);
          \draw[preaction={draw, line width=1.2pt, white},thick,blue] (0.5,1,0.7) to [out=-90, in=90] (1,0,0.7);
          \draw[preaction={draw, line width=1.2pt, white},thick,blue] (1,1,0.7) to [out=-90, in=90] (1.5,0,0.7);
          \draw[preaction={draw, line width=1.2pt, white},thick,blue] (1.5,1,0.7) to [out=-90, in=90] (0.5,0,0.7);  
              \filldraw[draw=black,fill=purple,thick] (0.57,0.75,0.7) circle (0.06);
              \filldraw[draw=black,fill=purple,thick] (1.05,0.75,0.7) circle (0.06);
               \filldraw[draw=black,fill=purple,thick] (1.4,0.75,0.7) circle (0.06);
               
              \draw[preaction={draw, line width=1.2pt, white},thick] (0.5,1,0.7)--(0.5,1,1.6)--(0.5,0,1.6)--(0.5,0,0.7);
              \draw[preaction={draw, line width=1.2pt, white},thick] (1,1,0.7)--(1,1,1.6)--(1,0,1.6)--(1,0,0.7);
                            \draw[preaction={draw, line width=1pt, white},thick] (1.5,1,0.7)--(1.5,1,1.6)--(1.5,0,1.6)--(1.5,0,0.7);
                       \pic at (1.5,0,0.7) {3dpepsp};
        \pic at (1.5,1,0.7) {3dpepsdownp};
         \pic at (1,0,0.7) {3dpepsp};
        \pic at (1,1,0.7) {3dpepsdownp};
        \pic at (0.5,0,0.7) {3dpepsp};
        \pic at (0.5,1,0.7) {3dpepsdownp};
                             \filldraw[draw=black,fill=orange,thick] (1.75,0,0.5) rectangle (1.75,0,0.9 );
            \filldraw[draw=black,fill=orange,thick] (0.25,1,0.5) rectangle (0.25,1,0.9);
                        \node[anchor= north east] at (0.5,0.25,1.6) {${b}$};
        \filldraw (0.5,0.25,1.6) circle (0.05);
           \draw[->] (0.5,0,0.7)--(0.5,0,1.3);
                                \node[anchor= north east] at (1,0.25,1.6) {${b}$};
        \filldraw (1,0.25,1.6) circle (0.05); 
           \draw[->] (1,0,0.7)--(1,0,1.3);
                                \node[anchor= north east] at (1.5,0.25,1.6) {${b}$};
        \filldraw (1.5,0.25,1.6) circle (0.05);   
           \draw[->] (1.5,0,0.7)--(1.5,0,1.3);      
             \node[anchor=west] at (2,0.7,0.7) {$\myinv{b}$};
        \filldraw (2,0.7,0.7) circle (0.05);
         \draw[->] (2,0.7,0.7) --(2,0.3,0.7);
 \node[anchor=east] at (0,0.3,0.7) {${b}$};
        \filldraw (0,0.3,0.7) circle (0.05);
         \draw[->] (0,0.5,0.7)--(0,0.7,0.7);           
\end{tikzpicture}
\times
\begin{tikzpicture}[scale=1]    
 \draw[preaction={draw, line width=1.2pt, white},thick] (1.8,0,0.7) rectangle (2.6,1,0.7);
 \draw[->] (1.8,1,0.7)--(2,1,0.7);
 \draw[->] (2.6,0,0.7)--(2.4,0,0.7);
              \filldraw[draw=black,fill=orange,thick] (2.2,1,0.5) rectangle (2.2,1,0.9 );  
            \filldraw[draw=black,fill=orange,thick] (2.2,0,0.5) rectangle (2.2,0,0.9  );
             \node[anchor=east] at (1.8,0.5,0.7) {${b}$};
        \filldraw (1.8,0.5,0.7) circle (0.05);
         \node[anchor=west] at (2.6,0.5,0.7) {$\myinv{b}$};
        \filldraw (2.6,0.5,0.7) circle (0.05);
   \end{tikzpicture}.
  \end{equation*}
Using Eq.\eqref{Ginjec} and Eq. \eqref{locsym}, we obtain a sum over $G=\mathbb{Z}_3$ for each of the three permuted sites (we denote these elements $s_1,s_2$ and $s_3$). The final expression to compute is the following
\begin{align}
\hat{\Lambda} &= \frac{1}{|G|^4}\sum_{s_1,s_2,s_3,b\in G} 
  \begin{tikzpicture}[baseline=+5mm]
        \draw[thick, blue] (0.6,0)..controls (1,0.5) and (2,0.5)..(2.4,1);
                \draw[thick, blue] (0.5,0)..controls (0.9,0.5) and (1.9,0.5)..(2.3,1);
                       \draw[preaction={draw, line width=1.2pt, white},thick,blue] (0.6,1) to [out=-90, in=90] (1.5,0);
        \draw[preaction={draw, line width=1.2pt, white},thick,blue] (0.5,1) to [out=-90, in=90] (1.4,0);
                       \draw[preaction={draw, line width=1.2pt, white},thick,blue] (0.6+0.9,1) to [out=-90, in=90] (1.5+0.9,0);
        \draw[preaction={draw, line width=1.2pt, white},thick,blue] (0.5+0.9,1) to [out=-90, in=90] (1.4+0.9,0);
                \draw[thick](0.6,0)--(1.4,0);
        \draw[thick](1.5,0)--(2.3,0);
        \draw[thick](2.4,0)--(2.8,0)--(2.8,1)--(2.4,1);
        \draw[thick](0.6,1)--(1.4,1);
        \draw[thick](1.5,1)--(2.3,1);
        \draw[thick](0.5,1)--(0,1);
        \draw[thick](0.5,0)--(0,0);
        \draw[thick](0,1)--(0,0);
        \node[anchor=south] at (0.35,1) {$\myinv{v_{q}}$};
        \filldraw[draw=black,fill=red,thick] (0.4,1) circle (0.06);
         \node[anchor=east] at (0.7,0.7) {$\myinv{s}_1$};
        \filldraw (0.58,0.75) circle (0.05);
        \node[anchor=east] at (1.5,0.7) {$\myinv{s}_2$};
        \filldraw (1.47,0.75) circle (0.05);
          \node[anchor=west] at (2.27,0.85) {${s_3}$};
        \filldraw (2.27,0.85) circle (0.05);
         \node[anchor=east] at (0.8,0.3) {$\myinv{s}_3$};
        \filldraw (0.76,0.25) circle (0.05);
        \node[anchor=west] at (2.33,0.3) {${s_2}$};
        \filldraw (2.33,0.25) circle (0.05);
        \node[anchor=west] at (1.4,0.2) {$s_1$};
        \filldraw (1.43,0.25) circle (0.05);
        \node[anchor=south] at (0.8,1) {$v_{q}$};
        \filldraw[draw=black,fill=red,thick] (0.8,1) circle (0.06);
        \draw[->] (0.9,1)--(1.1,1);
        \draw[->] (1.3,0)--(1.1,0);
        \draw[->] (1.3+0.8,0)--(1.1+0.8,0);
      \draw[->] (0.9+0.85,1)--(1.1+0.85,1);
         \node[anchor=south] at (1.3,1) {$\myinv{v_{q}}$};
        \filldraw[draw=black,fill=red,thick] (1.3,1) circle (0.06);
        \node[anchor=south] at (1.7,1) {$v_{q} $};
        \filldraw[draw=black,fill=red,thick] (1.7,1) circle (0.06);
                 \node[anchor=south] at (1.3+0.9,1) {$\myinv{v_{q}}$};
        \filldraw[draw=black,fill=red,thick] (1.3+0.8,1) circle (0.06);
        \node[anchor=south] at (2.6,1) {$v_{q} $};
        \filldraw[draw=black,fill=red,thick] (2.6,1) circle (0.06);
        \node[anchor=east] at (0,0.7) {$b$};
        \filldraw (0,0.7) circle (0.05);
        \draw[->] (0,0)--(0,0.4);
        \node[anchor=west] at (2.8,0.7) {$\myinv{b}$};
        \filldraw (2.8,0.7) circle (0.05);
        \draw[->] (2.8,1)--(2.8,0.4);
        \filldraw[draw=black,fill=orange,thick] (2.5,-0.1) rectangle (2.7,0.1);
        \filldraw[draw=black,fill=orange,thick] (0.05,0.9) rectangle (0.25,1.1);
         \end{tikzpicture}  \times 
 \notag \\
  &  \begin{tikzpicture}
       \draw[thick] (0,0) rectangle (1,1);
       \draw[->] (0,0.5)--(0,0.8);
 \draw[->] (1,0.5)--(1,0.2);
        \node[anchor=west] at (0,0.5) {$b$};
        \filldraw (0,0.5) circle (0.05);
       \node[anchor=east] at (1.1,0.6) {$\myinv{b}$};
        \filldraw (1,0.5) circle (0.05);
        \filldraw[draw=black,fill=orange,thick] (0.4,0.9) rectangle (0.6,1.1);
      \filldraw[draw=black,fill=orange,thick] (0.4,-0.1) rectangle (0.6,0.1);
     \end{tikzpicture}
  \times 
    \begin{tikzpicture}
          \draw[preaction={draw, line width=1.2pt, white},thick,blue] (0,0,-1) to [out=90, in=-90] (1.2,1,-1);
          \draw[preaction={draw, line width=1.2pt, white},thick,blue] (0.6,0,-1) to [out=90, in=-90] (0,1,-1);
 \draw[preaction={draw, line width=1.2pt, white},thick,blue] (1.2,0,-1) to [out=90, in=-90] (0.6,1,-1);
          \draw[->] (0,0.5,0)--(0,0.8,0);
          \draw[->] (0.6,0.4,0)--(0.6,0.6,0);
          \draw[->] (1.2,0,-1)--(1.2,0,-0.5);
        \foreach \x in {0,0.6,1.2}{
        \draw[preaction={draw, line width=1.2pt, white},thick] (\x,1,-1)--(\x,1,0)--(\x,0,0)--(\x,0,-1);
        \node[anchor= north east] at (\x,0.4,0) {$b$};
        \filldraw (\x,0.35,0) circle (0.05);
      \node[anchor=south] at (\x,1,-0.5) {$v_{q}$};
        \filldraw[draw=black,fill=red,thick] (\x,1,-0.5) circle (0.06);
        }
         \filldraw (0.15,0.3,-1) circle (0.05);
          \node[anchor=south east] at (0.2,0.25,-1) {$\myinv{s}_3$};
          \filldraw (0.5,0.3,-1) circle (0.05);
          \node[anchor=west] at (0.45,0.3,-1) {$\myinv{s}_1$};
          \filldraw (1.1,0.3,-1) circle (0.05);
          \node[anchor=west] at (1.05,0.3,-1) {$\myinv{s}_2$};
      \end{tikzpicture}
 \times  
        \begin{tikzpicture}
                        \draw[preaction={draw, line width=1.2pt, white},thick, blue] (0,0,1) to [out=90, in=-90] (1.2,1,1);
          \draw[preaction={draw, line width=1.2pt, white},thick,blue] (0.6,0,1) to [out=90, in=-90] (0,1,1);
 \draw[preaction={draw, line width=1.2pt, white},thick,blue] (1.2,0,1) to [out=90, in=-90] (0.6,1,1);
                  \draw[->] (0,0,0)--(0,0.3,0);
          \draw[->] (0.6,0.4,0)--(0.6,0.9,0);
          \draw[->] (1.2,0,0)--(1.2,0.3,0);
        \foreach \x in {0,0.6,1.2}{
        \draw[preaction={draw, line width=1.2pt, white},thick] (\x,1,1)--(\x,1,0)--(\x,0,0)--(\x,0,1);
        \node[anchor=  east] at (\x+0.15,0.6,0) {$\myinv{b}$};
        \filldraw (\x,0.5,0) circle (0.05);
      \node[anchor=south] at (\x,1,0.5) {$\myinv{v}_{q}$};
        \filldraw[draw=black,fill=red,thick] (\x,1,0.5) circle (0.06);
        }
         \filldraw (0.15,0.3,1) circle (0.05);
          \node[anchor=south east] at (0.2,0.25,1) {$s_3$};
          \filldraw (0.5,0.3,1) circle (0.05);
          \node[anchor=west] at (0.45,0.3,1) {$s_1$};
          \filldraw (1.1,0.3,1) circle (0.05);
          \node[anchor=west] at (1.05,0.3,1) {$s_2$};
 \end{tikzpicture}.\notag
 \end{align}
 Each term of this sum contains four diagrams. The first is made of two loops 
$$
\tr[   C^{[u]}_\sigma  \myinv{s}_1  s_3 \myinv{b} C^{[d]}_\sigma s_2 \myinv{s}_3 b ] \times \tr[ \myinv{s}_2 s_1].
$$
The last factor of this expression is equal to $|G|\delta_{s_2,s_1}$. The last two diagrams are both equal to $\tr[s_1s_2s_3v^{-3}_q b^{-3}]$ which reduces the sum to its terms that satisfy $v^3_q=s^2_1s_3$. Putting it all together, we get
$$
\hat{\Lambda} =\frac{1}{|G|}  \tr[ C^{[u]}_\sigma  v^3_q b^{-1} C^{[d]}_\sigma b v^3_q  ]\times \tr[  \bar{C}^{[u]}_\sigma  b^{-1} \bar{C}^{[d]}_\sigma b ]=\chi_{\sigma}( v^3_q).
$$
An analogous calculation can be carried out for arbitrary $p$ prime: we would also obtain a sum of four diagrams. The first contains $p-1$ loops:
$$\tr[   C^{[u]}_\sigma  \myinv{s}_1  s_p \myinv{b} C^{[d]}_\sigma s_2 \myinv{s}_p b ]  \tr[  s_1\myinv{s}_2] \tr[  s_2\myinv{s}_3]  \cdots \tr[  s_{p-2}\myinv{s}_{p-1}], $$
where the last $p-2$ factors reduces the sum to its elements that satisfy $s_1=s_2=\cdots =s_{p-2}=s_{p-1}$. The last two diagrams are both equal to $ \tr[s_1\cdots s_p v^{-p}_q b^{-p}] = \tr[s^{p-1}_1 s_p v^{-p}_q]$. So the only terms that survive in the sum are those satisfying $s_1s^{-1}_p=  v^p_q$. Finally
$$\hat{\Lambda }=  \chi_\sigma(v^p_q),$$ 
 where $\chi_\sigma$ denotes one of the $p-1$ non-trivial irreps of $\mathbb{Z}_p$. Therefore the order parameter is only equal to one if the {SF} pattern is trivial.

\subsection{ $\mathcal{D}$($\mathbb{Z}_4$) with $\mathbb{Z}_2$ symmetry }
       
We denote the topological group as $\mathbb{Z}_4=\{+1,-1,i,-i;\times \}$ and the symmetry group as $\mathbb{Z}_2=\{e,q;q^2=e \}$. There are two cases here to be analyzed depending on whether there is a non-trivial permutation action over the anyons or not. These permutations come from the possible homomorphism from $\mathbb{Z}_2$ to ${\rm Aut}(\mathbb{Z}_4)$. In each case there are two inequivalent {SF} classes.\\
 
 {\bf Non-trivial permutation}. In this case there is a non-trivial action of the symmetry operators over the topological group: $\phi_q(g)=v_q g v^{-1}_q= g^{-1}$;  $\forall g \in \mathbb{Z}_4$. That is, it permutes the fluxes $i$ and $-i$. There are two inequivalent cocycles since $H_\phi^2(\mathbb{Z}_2,\mathbb{Z}_4)= \mathbb{Z}_2$. 
If we denote the irreps of $\mathbb{Z}_4$ as $\sigma=0,1,2,3$, a non-trivial {SF} class (group extension $Q_8$) is characterized by a projective action on charges $1$ and $3$. Applying this action twice is equivalent to braiding with the flux $-1$. The symmetry permutes between charges $1$ and $3$, and multiplies the wave function of the system by sign factor.

The class can be distinguished with the quantity $\lambda=\omega(e,e)\omega(q,q)=v^2_{q}\in \mathbb{Z}_4$, which is represented by the identity element for the trivial {SF} class (group extension $D_8$) and $-1$ for the non-trivial one. $\lambda$ is not affected by gauge transformations because $v'^2_{q}= v^2_{q}l_q \phi_q(l_q)= v^2_{q}l_q l^{-1}_q=v^2_{q}$. This quantity measures if there are elements of the extension, that do not belong to $\mathbb{Z}_4$, with order greater than two.
We use the same configuration of operators as for the TC example, see Eq.(\ref{transprotocol}),  to construct the order parameter but creating the charges, irreps $\sigma=1,3$ of $\mathbb{Z}_4$, of the $\mathcal{D}$($\mathbb{Z}_4$). The final expression is the one of Eq.\eqref{loopcalc} particularising for the above relations of $G=\mathbb{Z}_4$. We get:
\begin{align*}
        \begin{tikzpicture}
        \draw[thick, blue] (1.5,1) to [out=-90, in=90] (0.6,0);
        \draw[thick,blue] (1.4,1) to [out=-90, in=90] (0.5,0);
       \draw[preaction={draw, line width=1.2pt, white},thick,blue] (0.6,1) to [out=-90, in=90] (1.5,0);
        \draw[preaction={draw, line width=1.2pt, white},thick,blue] (0.5,1) to [out=-90, in=90] (1.4,0);
        \draw[thick] (0.6,0)--(1.4,0);
        \draw[thick] (1.5,0)--(1.9,0);
        \draw[thick] (1.9,1)--(1.9,0);
        \draw[thick] (0.6,1)--(1.4,1);
        \draw[thick] (1.5,1)--(1.9,1);
        \draw[thick] (0.5,1)--(0,1);
        \draw[thick] (0.5,0)--(0,0);
        \draw[thick] (0,1)--(0,0);
        \node[anchor=south] at (0.35,1) {$\myinv{v_{q}}$};
        \filldraw[draw=black,fill=red,thick] (0.4,1) circle (0.05);
         \node[anchor=east] at (0.7,0.7) {$\myinv{a}$};
        \filldraw (0.58,0.75) circle (0.05);
        \node[anchor=west] at (1.4,0.7) {$c$};
        \filldraw (1.43,0.75) circle (0.05);
        \node[anchor=east] at (0.7,0.2) {$\myinv{c}$};
        \filldraw (0.58,0.25) circle (0.05);
        \node[anchor=west] at (1.4,0.2) {$a$};
        \filldraw (1.43,0.25) circle (0.05);
        \node[anchor=south] at (0.8,1) {$v_{q}$};
        \filldraw[draw=black,fill=red,thick] (0.8,1) circle (0.05);
         \draw[->] (1.2,0)--(1.0,0);
         \draw[->] (0.9,1)--(1.05,1);
         \node[anchor=south] at (1.3,1) {$\myinv{v_{q}}$};
        \filldraw[draw=black,fill=red,thick] (1.3,1) circle (0.05);
        \node[anchor=south,thick] at (1.9,1) {$v_{q} $};
        \filldraw[draw=black,fill=red] (1.7,1) circle (0.05);
        \node[anchor=west] at (0,0.5) {$b$};
        \draw[->] (0,0)--(0,0.3);
        \filldraw (0,0.5) circle (0.05);
        \node[anchor=west,thick] at (1.9,0.5) {$\myinv{b}$};
        \filldraw (1.9,0.5) circle (0.05);
        \draw[->] (1.9,0.9)--(1.9,0.7);
        \filldraw[draw=black,fill=orange,thick] (1.65,-0.1) rectangle (1.85,0.1);
        \filldraw[draw=black,fill=orange,thick] (0.05,0.9) rectangle (0.25,1.1);
         \end{tikzpicture}
     & = \tr[ v^{-1}_q c^{-1} b  C^{[u]}_\sigma v^{-1}_q a^{-1}  c v_q b^{-1} C^{[d]}_\sigma a v_q ]\\
     &= \tr[C^{[u]}_\sigma a c^{-1} b^{-1} C^{[d]}_\sigma a c^{-1} b ],
       \end{align*}
and the last two diagrams, see Eq. \eqref{loopcalc}, both equal to $\tr[ v_q a^{-1}b  v_q c^{-1} b]= \tr[ v^2_q a^{-1} c]$, are $|G|\delta_{ac^{-1}  ,v^{2}_q}$.
Next, we observe that $ \tr[C^{[u]}_\sigma a c^{-1} b^{-1} C^{[d]}_\sigma a c^{-1} b ]=\tr[ C^{[u]}_\sigma  v^2_q b^{-1} C^{[d]}_\sigma b v^2_q  ]$, which implies that together with the factor $\tr[  C^{[u]}_\sigma  b^{-1} C^{[d]}_\sigma b  ]$
$$
\hat{\Lambda} =\frac{1}{|G|}  \tr[ C^{[u]}_\sigma  v^2_q b^{-1} C^{[d]}_\sigma b v^2_q  ]\times \tr[  \bar{C}^{[u]}_\sigma  b^{-1} \bar{C}^{[d]}_\sigma b ]=\chi_{\sigma}( v^2_q).
$$

This calculation is valid for any $\sigma$ but to distinguish between phases, we have to choose the two irreps, $\sigma=1,3$, that satisfy $\chi_{\sigma}( -1)\neq1$ (the ones that fractionalize the symmetry).\\

 {\bf Trivial permutation}. 
 There are two fractionalization classes since $H^2(\mathbb{Z}_2,\mathbb{Z}_4)= \mathbb{Z}_2$. A gauge-invariant quantity is $\lambda=v^4_q=\omega^2(e,q) \omega^2(q,q)$ since $v'^4_q=v^4_ql^4_q$ and $l_q\in \mathbb{Z}_4$. The value of $\lambda$ is $-1\in \mathbb{Z}_4$ for the non-trivial class (group extension $\mathbb{Z}_8$) and the identity for the trivial one (group extension $\mathbb{Z}_4\times \mathbb{Z}_2$). $\lambda$ measures whether there are elements of order greater than $4$ in the group extension. The order parameter is
   \begin{equation} 
   \Lambda=\begin{tikzpicture}[scale=1]
         \pic at (1,0,0) {3dpeps};
        \pic at (1,1,0) {3dpepsdown};
        \pic at (0.5,0,0) {3dpeps};
        \pic at (0.5,1,0) {3dpepsdown};
        \pic at (1.5,0,0) {3dpeps};
        \pic at (1.5,1,0) {3dpepsdown};
          \pic at (2,0,0) {3dpeps};
        \pic at (2,1,0) {3dpepsdown};
                 \filldraw[draw=black,fill=orange,thick] (0.75,1,-0.2) rectangle (0.75,1,0.2);  
          \filldraw[draw=black,fill=orange,thick] (0.75,0,-0.2) rectangle (0.75,0,0.2 );
                  \begin{scope}[canvas is xy plane at z=0]
          \draw[preaction={draw, line width=1.2pt, white},thick,blue] (2,1,0.7) to [out=-90, in=90] (0.5,0,0.7);
          \draw[preaction={draw, line width=1.2pt, white},thick,blue] (0.5,1,0.7) to [out=-90, in=90] (1,0,0.7);
          \draw[preaction={draw, line width=1.2pt, white},thick,blue] (1,1,0.7) to [out=-90, in=90] (1.5,0,0.7);
          \draw[preaction={draw, line width=1.2pt, white},thick,blue] (1.5,1,0.7) to [out=-90, in=90] (2,0,0.7);
        \end{scope}
        \filldraw[draw=black,fill=purple,thick] (1.97,0.85,0.7) circle (0.07);
                 \filldraw[draw=black,fill=purple,thick] (1.53,0.85,0.7) circle (0.07);
         \filldraw[draw=black,fill=purple,thick] (1.03,0.85,0.7) circle (0.07);
         \filldraw[draw=black,fill=purple,thick] (0.53,0.85,0.7) circle (0.07);
                 \pic at (0,0,0.7) {3dpeps};
        \pic at (0,1,0.7) {3dpepsdown};
        \pic at (0.5,0,0.7) {3dpepsp};
        \pic at (0.5,1,0.7) {3dpepsdownp};
        \pic at (1,0,0.7) {3dpepsp};
        \pic at (1,1,0.7) {3dpepsdownp};
           \pic at (1.5,0,0.7) {3dpepsp};
        \pic at (1.5,1,0.7) {3dpepsdownp};
                \pic at (2,0,0.7) {3dpepsp};
        \pic at (2,1,0.7) {3dpepsdownp};
                   \pic at (2.5,0,0.7) {3dpeps};
        \pic at (2.5,1,0.7) {3dpepsdown};
         \pic at (1,0,1.4) {3dpeps};
         \pic at (0.5,0,1.4) {3dpeps};
        \pic at (0.5,1,1.4) {3dpepsdown};
         \pic at (1.5,0,1.4) {3dpeps};
         \pic at (2,0,1.4) {3dpeps};      
        \draw[thick] (1,1,1.4)--(1,0.81,1.4);
    \begin{scope}[canvas is zx plane at y=1]
      \draw[preaction={draw, line width=1.2pt, white},thick] (0.9,1)--(1.9,1);
      \draw[preaction={draw, line width=1.2pt, white},thick] (1.4,0.6)--(1.4,1.4);
      \filldraw (1.4,1) circle (0.07);
    \end{scope} 
            \draw[thick] (1.5,1,1.4)--(1.5,0.81,1.4);
    \begin{scope}[canvas is zx plane at y=1]
      \draw[preaction={draw, line width=1pt, white},thick] (0.9,1.5)--(1.9,1.5);
      \draw[preaction={draw, line width=1pt, white},thick] (1.4,1.1)--(1.4,1.9);
      \filldraw (1.4,1.5) circle (0.07);
    \end{scope} 
             \draw[thick] (2,1,1.4)--(2,0.81,1.4);
    \begin{scope}[canvas is zx plane at y=1]
      \draw[preaction={draw, line width=1.2pt, white},thick] (0.9,2)--(1.9,2);
      \draw[preaction={draw, line width=1.2pt, white},thick] (1.4,1.6)--(1.4,2.4);
      \filldraw (1.4,2) circle (0.07);
    \end{scope} 
      \filldraw[draw=black,fill=orange,thick] (2.25,0,0.5) rectangle (2.25,0,0.9);
       \filldraw[draw=black,fill=orange,thick] (0.25,1,0.5) rectangle (0.25,1,0.9); 
 \end{tikzpicture} \notag.\end{equation}
The calculation done in Section \ref{secq} (for $p=4$) is also valid here, since $\phi_q=\id$, therefore  
$$\hat{\Lambda} =  \chi_{\sigma}( v^4_q),$$
where  $\sigma$ denotes  the two irreps of $\mathbb{Z}_4$ that satisfy $\chi_{\sigma}( -1)\neq1$.

\subsection{ $\mathcal{D}(Q_8)$ with $\mathbb{Z}_2$ symmetry}\label{ex:nonab}

Here we study a case of non-abelian topological order: $\mathcal{D}$($Q_8$) with $Q=\mathbb{Z}_2=\{e,q;q^2=e \}$ symmetry and trivial anyon permutation action which can host two inequivalent {SF} classes. The non-trivial class is associated with a projective action of $\mathbb{Z}_2$ over the two-dimensional charge (irrep). The two related group extensions are $\mathbb{Z}_2 \times Q_8 \equiv \{ (s,g)| s=0,1 ; g\in Q_8 \}$ and $(\mathbb{Z}_4 \times Q_8)/\mathbb{Z}_2 \equiv \{ (t,g)| t=0,1,2,3 ; g\in Q_8 \} / \{(0,e),(2,-1)  \}$.
Let us consider the posible values of $v_q$ and $v^2_q=\omega(e,e)\omega(q,q)\in Q_8 $ in both group extensions:

\begin{table}[htb]
\centering
\begin{tabular}{|c|c|}
\hline
\multicolumn{2}{|c|}{$\mathbb{Z}_2 \times Q_8$} \\
 \hline\hline
$v_q$ & $v^2_q\in Q_8$ \\ \hline 
$(1,+1)$ & $+1$ \\ \hline
$(1,-1)$ & $+1$ \\ \hline
$(1,+i)$ & $-1$ \\ \hline
$(1,-i)$ & $-1$ \\ \hline
$(1,+j)$ & $-1$ \\ \hline
$(1,-j)$ & $-1$ \\ \hline
$(1,+k)$ & $-1$ \\ \hline
$(1,-k)$ & $-1$ \\ \hline
\end{tabular}
\qquad
\begin{tabular}{|c|c|}
\hline
\multicolumn{2}{|c|}{$(\mathbb{Z}_4 \times Q_8)/\mathbb{Z}_2 $} \\
 \hline\hline
$v_q$ & $v^2_q\in Q_8$ \\ \hline 
$(1,+1)\equiv (3,-1)$ & $-1$ \\ \hline
$(1,-1)\equiv (3,+1)$ & $-1$ \\ \hline
$(1,+i)\equiv (3,-i)$ & $+1$ \\ \hline
$(1,-i)\equiv (3,+i)$ & $+1$ \\ \hline
$(1,+j)\equiv (3,-j)$ & $+1$ \\ \hline
$(1,-j)\equiv (3,+j)$ & $+1$ \\ \hline
$(1,+k)\equiv (3,-k)$ & $+1$ \\ \hline
$(1,-k)\equiv (3,+k)$ & $+1$ \\ \hline
\end{tabular}
\end{table}
The fact that the topological group is non-abelian implies that the natural candidate for $\lambda$ of a $\mathbb{Z}_2$ symmetry, $v^2_q$, is not gauge-invariant. Besides that, there exists a difference in $v^2_q$ between extensions in the number of times that it can be $-1$ (or $+1$). This distinction can be captured by the magnitude:
\begin{align*}
\lambda&= \sum_{g\in Q_8} (v_q\times g)^2 =  \sum_{g\in Q_8}(1,g)^2 \\
 &=\bigg\{ \begin{array}{lcc}   
 6u_{+1}+2u_{-1}& {\rm in} & (\mathbb{Z}_4 \times Q_8)/\mathbb{Z}_2 
\\
  6u_{-1}+2u_{+1} &  {\rm in} &\mathbb{Z}_2 \times Q_8 \end{array},
\end{align*}
which does not depend on the gauge and where $u_{+1},u_{-1}$ correspond to the representation of $Q_8$ of the elements $+1,-1$.  Let us notice that $\lambda$ here does not belong to $Q_8$, but it belongs to the algebra generated by the representation of $G$ acting on the virtual d.o.f. This suggests that $\lambda$ can be interpreted as a superposition of fluxes which depends on the extension. To obtain $\lambda$ at the virtual level, we use again the same configuration of operators as for the TC example, see Eq.(\ref{transprotocol}),  to construct the order parameter. But we now create the non-abelian charge of  $\mathcal{D}$($Q_8$) which corresponds to the two-dimensional irrep of $Q_8$.
Then, the final expression is the one of Eq.\eqref{loopcalc} particularising it for this case. There are two factors equal to $\tr[c v^{-1}_q b^{-1} a v^{-1}_q b^{-1}]$ which reduce the sum over $c$ in Eq.\eqref{loopcalc} to the terms where $c^{-1}= v^{-1}_qb^{-1}av^{-1}_q b^{-1}$. This implies that 
 $$ac^{-1}b =(av^{-1}_qb^{-1})(av^{-1}_q b^{-1})b= [v^{-1}_q a^{-1}  c v_q b^{-1}]^{-1}.$$
The factor $av^{-1}_q b^{-1}$ can be relabelled as $a\phi_q(b^{-1})v^{-1}_q\equiv g^{-1} v^{-1}_q$ with $g\in Q_8$  replacing the sum over $a$ and $b$ with the sum over $g$ (and a factor $|Q_8|$). The remaining two factors are $\tr[  \bar{C}^{[u]}_{\sigma,h}  b^{-1} \bar{C}^{[d]}_{\sigma,h} b ]$ and $\tr[ v^{-1}_q c^{-1} b  C^{[u]}_\sigma v^{-1}_q a^{-1}  c v_q b^{-1} C^{[d]}_{\sigma,h} a v_q ]= \tr[ C^{[u]}_{\sigma,h} v^{-1}_q a^{-1}  c v_q b^{-1} C^{[d]}_{\sigma,h} ac^{-1}b ]$ where in the last relation we can identify the factor $ac^{-1}b$ previously discussed. We finally obtain:

\begin{align}
\hat{\Lambda} & \propto  \sum_{b,g,h\in Q_8} \tr[ C^{[u]}_{\sigma, h}  b^{-1}(v_q g)^{2} C^{[d]}_{\sigma , h}  (v_q g )^{-2} b ] \tr[  \bar{C}^{[u]}_{\sigma , h}  b^{-1} \bar{C}^{[d]}_{\sigma ,h} b ] \notag \\
& =  \bigg\{ \begin{array}{ccc}  
\left(6\chi_\sigma(+1)+2\chi_\sigma(-1)\right)/8& {\rm for} & E = (\mathbb{Z}_4 \times Q_8)/ \mathbb{Z}_2\\ 
\left( 6 \chi_\sigma(-1) +2\chi_\sigma(+1) \right) /8 &  {\rm for} & E = \mathbb{Z}_2 \times Q_8 
		\end{array}\notag \\
 & = \bigg\{ \begin{array}{lcc} 
  			+1 & {\rm for } &  E = (\mathbb{Z}_4 \times Q_8)/\mathbb{Z}_2  \\
  			-1 &  {\rm for } & E = \mathbb{Z}_2 \times Q_8,
   		\end{array}\notag
\end{align}
where in the last step we have used that $\sigma$ is the two-dimensional irrep of $Q_8$. The fact that $v^2_q$ belongs to the group algebra instead of just to the group comes from $Q_8$ being non-abelian. 

\section{Conclusions and outlook}

In this paper, we have presented order parameters to distinguish the fractionalization class of an internal symmetry in $G$-isometric {PEPS}. These are calculated in the bulk of the {2D} system and only depend on the virtual symmetry operators. Our technique works for some very interesting cases like non-abelian topological order or {SF} classes involving permutations of anyons. 
We have also shown in all the examples provided that the {SF} classes are better resolved with our order parameters than would be possible through dimensional compactification.

The calculations have been done using square lattices but they can be generalized. For example in the hexagonal lattice if the corresponding trivalent tensors are $G$-injective, blocking two of them we end up with a square lattice and the {SF} class could be identified. We notice that this result can be potentially useful for a generalisation of our order parameter in general string nets models\cite{Levin05} enriched with symmetries. To do so it would be interesting to study how our findings could be adpated to the case where the symmetry is not internal and when it is not represented virtually as a tensor product, but as a matrix product operator (MPO) \cite{Williamson17}. In fact the MPO-like form would allow to realize more SF patterns: in our approach we only consider SF patterns equivalent to braiding with fluxes.

The construction of order parameters for non-unitaries and lattice symmetries is left for future work; some gauge-invariant quantitites of these symmetries are calculated in Appendix \ref{apendixTRS}.

Our order parameter is defined for the RG fixed points of $G$-injective {PEPS}, $G$-isometric PEPS, where the correlation length is zero and Eq.(\ref{Gisoconca}) can be used to compute analytically all the expressions.  At those points we can ensure that the state is in the $\mathcal{D}$($G$) phase in the thermodynamic limit.
RG procedures in 2D tensor network states are not trivial, see \cite{Hauru} and references therein. However, for $G$-injective PEPS that converge 
to a $G$-isometric PEPS when blocking, Eq.(\ref{Gisoconca}) is valid. We expect that our order parameters work for those tensor networks and RG procedures because the maps $(\omega, \phi)$ are well defined in arbitrary $G$-injective PEPS and are left invariant under blocking. It can be shown that the blocking version of our order parameters work for RFP. For example the blocked version of \eqref{transprotocol}:
   \begin{equation*} 
\Lambda= \begin{tikzpicture}[baseline=+1mm][scale=1]
  \pic at (0,0,0.7) {3dpeps};
   \draw[thick](0,0,0.7)--(0,1,0.7);
        \pic at (0,1,0.7) {3dpepsdown};
            \pic at (4.4,0,0.7) {3dpeps};
             \draw[thick](4.4,0,0.7)--(4.4,1,0.7);
       \pic at (4.4,1,0.7) {3dpepsdown};  
        \draw[thick](0,1,0.7)--(4.4,1,0.7);
        \draw[thick](0,0,0.7)--(4.4,0,0.7);
          \begin{scope}[canvas is xy plane at z=0]
            \foreach \x in {0.5,0.7,0.9,1.1,1.3}{
                      \draw[preaction={draw, line width=1.2pt, white},thick, blue] (\x,1,0.7) to [out=-90, in=90] (1.3+\x,0,0.7);
            }  
              \foreach \x in {0.5,0.7,0.9,1.1,1.3}{
                    \draw[preaction={draw, line width=1.2pt, white},thick,blue] (\x+1.3,1,0.7) to [out=-90, in=90] (\x,0,0.7);
                    \filldraw[draw=black,fill=purple,thick] (\x+0.05,0.8,0.7) circle (0.06);
                     \filldraw[draw=black,fill=purple,thick] (\x+1.3-0.05,0.8,0.7) circle (0.06);
            }  
                  \foreach \x in {0.5,0.7,0.9,1.1,1.3}{
                            \pic at (\x,0,0.7) {3dpepspshort};
    			   \pic at (\x,1,0.7) {3dpepsdownpshort};
                             \pic at (1.3+\x,0,0.7) {3dpepspshort};
     			 \pic at (1.3+\x,1,0.7) {3dpepsdownpshort};
			  \draw[thick](2.6+\x,0,0.7)--(2.6+\x,1,0.7);
			 \pic at (2.6+\x,0,0.7) {3dpepspb};
     			 \pic at (2.6+\x,1,0.7) {3dpepsdownpb};
    } 
       \end{scope}
       \draw [decorate,decoration={brace,raise=5pt},yshift=0pt]
(0.4,1,0.7) -- (1.4,1,0.7) node [black,midway,yshift=0.4cm] {\footnotesize
$r$};
             \filldraw[draw=black,fill=orange,thick] (2.85,0,0.5) rectangle (2.85,0,0.9 );
            \filldraw[draw=black,fill=orange,thick] (0.25,1,0.5) rectangle (0.25,1,0.9); 
  	     \filldraw[draw=black,fill=orange,thick] (4.15,0,0.5) rectangle (4.15,0,0.9 );
  	     \filldraw[draw=black,fill=orange,thick] (4.15,1,0.5) rectangle (4.15,1,0.9 );
\end{tikzpicture},
\end{equation*}
where $r$ should be taken larger than the correlation length. 
 These considerations are important for the future numerical implementation of our order parameters and will be investigated in details in a future work.
We notice that the locality of our approach allows to use 2D techniques as the Corner Transfer Method (CTM) -see \cite{Nishino96} and \cite{Fishman18} for a recent review.

We also point out that the locality of our order parameters doesn't contradict the fact that topological order cannot be detected locally. 
But remarkably the SF pattern is calculated thanks to its duality with braiding, so our detection implies a non-trivial braiding of the excitations of the
quantum phases that we are considering (besides in our calculations only one type of anyon is considered).\\

\section*{Aknowledgements}

J.G.R. acknowledges Andras Molnar for the introduction to Ti$k$Z and for the examples provided. The authors acknowledge financial support from MINECO (grant MTM2014- 54240-P), from Comunidad de Madrid (grant QUITEMAD+- CM, ref. S2013/ICE-2801), and the European Research Council (ERC) under the European Union's Horizon 2020 research and innovation programme (grant agreement No 648913). S.I. acknowledges financial support from Ministerio de Ciencias, Innovaci\'on y Universidades, "Simulaci\'on y computaci\'on cu\'anticas", PGC2018-095862-B-C22. This work has been partially supported by ICMAT Severo Ochoa project SEV-2015-0554 (MINECO).

\appendix  

\section{Group extensions}\label{ap:ext}

An extension of a group $Q$ is a group $E$, together with a surjective homomorphism $\pi:E\rightarrow Q$. The kernel of  $\pi$ is a normal subgroup $G$ of $E$. We say that the group $E$ with the homomorphism $\pi$ is an extension of $Q$ by $G$ \cite{Adem04}. An extension is encoded in the following short exact sequence:
$$1\rightarrow  G  \stackrel{i}{\rightarrow} E\stackrel{\pi}{\rightarrow} Q \rightarrow 1,$$
where $i$ is the inclusion map and $Q$ is isomorphic to the quotient group $E/G$.\\

In the case where $G$ is an abelian group, given a group extension $E$ of $Q$ by $G$ and the homomorphism $\pi$, two maps can be defined: (i) a homomorphism $\phi:Q  \rightarrow  {\rm Aut}(G)$ and (ii) a $2$-cocycle $\omega:Q\times Q\rightarrow G$ which satisfies 
\begin{equation} \label{concocycle}
\omega(k,q)\omega(kq,p)=\phi_k(\omega(q,p))\omega(k,qp).
\end{equation}
These maps are defined as follows. Given $k$, we pick a pre-image $\epsilon_k\in E$ such that $\pi(\epsilon_k)=k$, and we construct $\phi_k:g\mapsto \epsilon_k g \epsilon^{-1}_k$ and $\omega(k,q)=\epsilon_k \epsilon_q \epsilon^{-1}_{kq}$. There is some arbitrariness in the choice of the pre-image: $\epsilon_k$ and $l_k\epsilon_k $ maps to $k$ under $\pi$ for any $l_k \in G$.  This  freedom in the choice of the pre-image does not modify the map $\phi_k$, but it affects the cocycle as follows
\begin{equation} \label{relcocycle}
\omega'(k,q)= l_k \phi_k(l_q)l^{-1}_{kq}\omega(k,q),
\end{equation}
for $k,q\in Q$. The second cohomology group $H_\phi^2(Q,G)$ is defined as the $2$-cocycles satisfying Eq.(\ref{concocycle}) and quotienting by $2$-coboundaries, \emph{i.e.} maps of the form $ Q \times Q \to G: ( k,q) \to l_k \phi_k(l_q)l^{-1}_{kq}$. That is, we identify cocycles that are related as Eq.(\ref{relcocycle}).

Two extensions are said to be equivalent if there is an isomorphism $\sigma:E\rightarrow E'$ such that the following diagram commutes:
\beq \label{eq:extdia}
\begin{array}{lllllllll}
1 & \longrightarrow  &G &\stackrel{i}{\longrightarrow} & E& \stackrel{\pi}{\longrightarrow}
& Q& \longrightarrow  & 1 \\
&  & \downarrow \,\id &  & \downarrow \,\sigma &  & \downarrow \,\id &  &  \\
1 & \longrightarrow  & G & \stackrel{i'}{\longrightarrow} &E' & \stackrel{\pi'}{\longrightarrow}
& Q & \longrightarrow  & 1.
\end{array}
\eeq
If $E$ and $E'$ come from a commutative diagram as Eq. (\ref{eq:extdia}) then, they are isomorphic as groups. However it is possible that the diagram (\ref{eq:extdia}) does not commute even though $E\equiv E'$ constructed with the same groups $Q$ and $G$. Hence equivalence of extensions is a more subtle notion than group equivalence. An important result is that if two extensions are equivalent then the action $\phi:Q\rightarrow {\rm Aut}(G)$ is the same for both, and the 2-cocycles describing the two extensions are the same in $H_\phi^2(Q,G)$.\\

To deal with the non-abelian case, two maps $\omega$ and $\phi$ can also be constructed as in the abelian case. But the map $\phi:Q\rightarrow {\rm Aut}(G)$ need not be a group homomorphism now. In fact it satisfies
$$\phi_k \circ \phi_q={\rm Inn}(\omega(k,q)) \circ \phi_{kq},$$
where Inn$(g)$ denotes the inner automorphism $h\mapsto g h g^{-1}: g,h \in G$. The map $\omega(k,q)$ is defined as in the abelian case and also satisfies the cocycle condition (\ref{concocycle}). However, the group homomorphism $\phi$ now maps elements of $Q$ to $ {\rm Out}(G) \equiv {\rm Aut}(G)/{\rm Inn}(G)$. The extension group equivalence is again defined as the commutation of the diagram (\ref{eq:extdia})  and is classified by $\phi$ and the cocycle $\omega$. It can be shown that the group $H_\phi^2(Q,Z(G))$ acts freely and transitively over the set of extensions, where $Z(G)$ is the center of $G$ \cite{Adem04}. In particular, this implies that $|H_\phi^2(Q,Z(G))|$ is equal to the number of the inequivalent cocycles. The elements of $H_\phi^2(Q,Z(G))$ are constructed as $c(q,k)=\omega'(q,k)\omega^{-1}(q,k)$, \textit{i.e.} the difference between cocycles, so that the non-trivial element maps one class into another. That is, the difference between  cocycles of non-abelian groups is given by a second cohomology group that classifies the general cocycles of the abelian groups. 

\section{Projective representations}\label{projrep}

A projective representation of a group $Q$ is a homomorphism from $Q$ to $GL(n)$ up to a phase factor:
$$
V_qV_k=\rho(q,k) V_{qk};\; \forall q,k\in Q,
$$
where $\rho(q,k)\in U(1)$. Associativity of group multiplication implies the so called cocycle condition: $ \rho(q,k) \rho(qk,p)=  \rho(k,p)  \rho(q,kp)$. A change of basis of the vector space where $\{V_q\}$ act does not affect $\{\rho(q,k)\}$ but a phase redefinition $V_q\to V'_q \equiv \nu_q V_q$ induces the modification 
\beq\label{eq:equiv-rho}
\rho(q,k) \to \rho'(q,k) \equiv \nu^{-1}_q \nu^{-1}_k \nu_{qk}\rho(q,k).
\eeq
Eq.\eqref{eq:equiv-rho} is taken to be the equivalent relation to classify cocycles resulting in the group $H^2(Q,U(1))$.  An example of projective representation of $\mathbb{Z}_2\times \mathbb{Z}_2=\{x^2=y^2=z^2=e, xy=z \}$ is given by the Pauli matrices.
\

Given $Q$ there exists a group $E$ such that a projective representation $V$ of $Q$ can be expressed as a linear representation $U$ of $E$. $E$ is a so-called representation group of $G$ \cite{Schur04}. The relation of the two groups is the extension $1\rightarrow G \rightarrow E \rightarrow Q \rightarrow1$ where $G$ is a group that satisfies $U_g\propto \id;\;\forall g \in G$. For the previous example $E=Q_8=\{ \pm 1,\pm i,\pm j,\pm k\}$ with the representation $U_i=i\sigma_z$, $U_j=-i\sigma_y$, $U_k=-i\sigma_x$.

\

Considering the definition of $\omega$ in the previous section
$$
\epsilon_k \epsilon_q =\omega(k,q) \epsilon_{kq},
$$
one can take a faithful representation $W$ of $E$ to realize this equation. The representation $V_q\equiv W_{\epsilon_q}$ can be seen as a \emph{projective} representation of $Q$:
$$
V_k V_q =W_{\omega(k,q)} V_{kq},
$$
where the projective factors are matrices: the representation $W$ restricted to the elements of $G\subset E$.
For example the case $G=Q=\mathbb{Z}_2$ realizing $\omega(-1,-1)=-1$ is $E=\mathbb{Z}_4$ where one can take the following representation:
$$
V_{-1}=
  \left( {\begin{array}{cccc}
   0 & 1 & 0&0 \\
 0 & 0 & 1&0 \\
  0 & 0 & 0&1 \\
   1 & 0 & 0&0 
  \end{array} } \right), \;
  W_{-1}=
  \left( {\begin{array}{cccc}
   0 & 0 & 1&0 \\
 0 & 0 & 0&1 \\
  1 & 0 & 0&0 \\
   0 & 1 & 0&0 
  \end{array} } \right).
  $$
These matrices allow us to construct the non-trivial phase corresponding to the example of \cref{TCZ2} (see \cite{Garre17}). The tensor is given by:
 $$A=\frac{1}{4} \sum_{b=0,1} W^b_{-1}\otimes W^b_{-1}\otimes W^{-b}_{-1}\otimes W^{-b}_{-1},$$
the physical operator of the global symmetry is
$$U_{-1}=V_{-1}\otimes V_{-1}\otimes V^{-1}_{-1}\otimes V^{-1}_{-1},$$
and the virtual representation of the charge operator is
$$C_\sigma= \sigma_z \otimes \id_2.$$
Altogether they satisfy that the projective action of the symmetry over the charge is equivalent to the braiding with the flux:
$$ V^2_{-1} C_\sigma V^{-2}_{-1}= W_{-1} C_\sigma W^{-1}_{-1}=-C_\sigma.$$

\section{Analysis of $H^2(\mathbb{Z}_2 \times \mathbb{Z}_2 ,\mathbb{Z}_2)$ }\label{CohomoTCZ2Z2}

In this appendix we show how the group $H^2(\mathbb{Z}_2 \times \mathbb{Z}_2 ,\mathbb{Z}_2)=\mathbb{Z}_2 \times\mathbb{Z}_2 \times\mathbb{Z}_2 \equiv \mathbb{Z}^3_2$ is reduced to four classes, related to $\{ \mathbb{Z}^3_2, \mathbb{Z}_4\times \mathbb{Z}_2, D_8,Q_8\}$, after taking the relabelling in the group $\mathbb{Z}_2 \times \mathbb{Z}_2=\{e,x,y,xy\}$ into account. 

\begin{figure}[ht!]
\begin{center}
\includegraphics[scale=0.8]{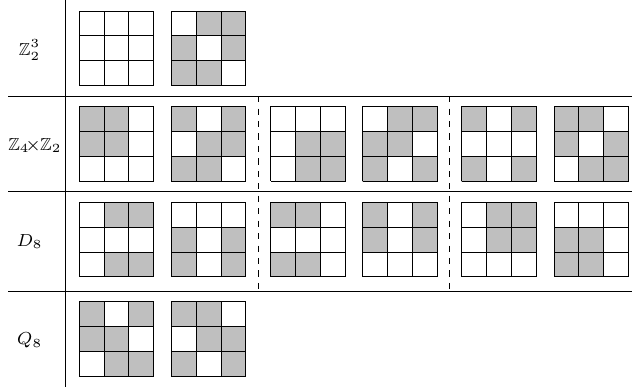}
\caption{The different cocycles of $\mathbb{Z}_2\times \mathbb{Z}_2$ by $\mathbb{Z}_2$ related to the extensions. The dashed lines separate the inequivalent cocycles coming from the classification of the second cohomology group; these are related by an automorphism of $\mathbb{Z}_2\times \mathbb{Z}_2$ giving the same group.}
\label{fig:cocyclesQ8Z2}
\end{center}
\end{figure}

We can always choose that $\omega(e,q)=\omega(q,e)=1$ for any $q\in \mathbb{Z}_2 \times \mathbb{Z}_2$   so we only write the elements $\omega(q,k)$ where $q$ and $k$ are both different from $e$. Thus we represent the different 2-cocycles as a $3\times 3$ table where white indicates $+1$ and grey $-1$. The trivial 2-cocycles, 2-coboundaries, can be constructed as $\omega(q,k)=g_q g_k g^{-1}_{qk}$ where $g: \mathbb{Z}_2 \times \mathbb{Z}_2 \to \mathbb{Z}_2$.  There are two distinct 2-coboundaries shown in Fig. \ref{fig:cocyclesQ8Z2}. These 2-coboundaries are the gauge freedom of the 2-cocycles in the second cohomology group and correspond to the trivial extension, this is the direct product of $\mathbb{Z}_2 \times \mathbb{Z}_2$ and $\mathbb{Z}_2$. We note that the quantities $\lambda=\{\omega(q,q)\}$ and $\lambda'=\omega(q,k)\omega(k,q)^{-1}$ for $q,k\in \mathbb{Z}_2\times \mathbb{Z}_2$ are invariant under this gauge freedom. Given an automorphism $\alpha$ of $\mathbb{Z}_2\times \mathbb{Z}_2$, we group together 2-cocycles $\omega, \omega'$ related by $\omega'(q,k)=\omega(\alpha(q),\alpha(k))$. It is also clear that this  composition does not change the value of $\lambda$ and $\lambda'$; these quantities are invariant under all the allowed freedom. The possible 2-cocycles are shown in Fig. \ref{fig:cocyclesQ8Z2}.       
This example shows that the classification of these phases loses the group structure of the second cohomology group. 
      
 \section{Time reversal symmetry}\label{apendixTRS}

Let us consider a $G$-injective {PEPS} invariant under time reversal symmetry $ \mathcal{T} |\psi\rangle=  |\psi\rangle$ which is realised by an anti-unitary global operator $\mathcal{T}=U_{\mathcal{T}}^{\otimes N}K$, where $U_{\mathcal{T}}$ is unitary and $K$ is the complex conjugation operator and we will denote is action as $K v= v^*$ where $^*$ means complex conjugation. The local transformation of the tensors is
\begin{equation}\label{eq:TRS}
U_{\mathcal{T}}A^*=A(\myinv{V}_{\mathcal{T}}\otimes V_{\mathcal{T}}\otimes \myinv{V}_{\mathcal{T}}\otimes V_{\mathcal{T}}).
\end{equation}
The condition $ \mathcal{T}^2=\id$ implies $A= U_{\mathcal{T}}(U_{\mathcal{T}}A^*)^*= A ({\myinv{V}_{\mathcal{T}}}^* \myinv{V}_{\mathcal{T}}\otimes V_{\mathcal{T}}V^*_{\mathcal{T}}\otimes {\myinv{V}_{\mathcal{T}}}^* \myinv{V}_{\mathcal{T}}\otimes V_{\mathcal{T}}V^*_{\mathcal{T}})$
\cite{Molnarinprep}. Therefore, we conclude that $$V_{\mathcal{T}}V^*_{\mathcal{T}}\equiv\omega_{\mathcal{T}}\in G.$$ We point out that $A^*$ is a $G$-injective tensor with respect to the conjugated representation of $G$ acting on $A$ that we will denote as $g^*\in G^*$. Using Eq.(\ref{eq:TRS}) and the corresponding $G$-injectivity of both $A^*$ and $A$,  the following holds $$V_{\mathcal{T}}g^*\myinv{V}_{\mathcal{T}}\equiv\phi_{\mathcal{T}}(g^*)\in G\;\Rightarrow \phi_{\mathcal{T}}\circ \phi^*_{\mathcal{T}}=\tau_{\omega_{\mathcal{T}}},$$ where $\phi^*_{\mathcal{T}}(g)\equiv V^*_{\mathcal{T}}g{\myinv{V}_{\mathcal{T}}}^*\in G^*$. We notice the difference with an  internal $\mathbb{Z}_2=\{1,k\}$ symmetry where we would have obtained $V^2_k\in G$ and $V_k g \myinv{V}_k\in G$. If the representation of $G$, in some basis, is real then $U_{\mathcal{T}}A=A(\myinv{V}_{\mathcal{T}}\otimes V_{\mathcal{T}}\otimes \myinv{V}_{\mathcal{T}}\otimes V_{\mathcal{T}})$. This is the case for $G$-isometric {PEPS} with the left regular representation in the group algebra basis: $L_g=\sum_{h\in G}|gh\rangle\langle h|\Rightarrow  L^*_g=L_g$ which implies that $ A=A^*$.
From Eq.(\ref{eq:TRS}), it is clear that the operator $V_{\mathcal{T}}$ is defined up to an element of $G$: $V_{\mathcal{T}}\sim gV_{\mathcal{T}}$ so%
$$ \omega'_{\mathcal{T}} = g \phi_{\mathcal{T}} (g^*) \omega_{\mathcal{T}}. $$%
We can define recursively the coefficient of the $m$-power of $\omega'_{\mathcal{T}}$: $${\omega'_{\mathcal{T}}}^m = h_m \omega_{\mathcal{T}}^m,$$ where $ h_m=h_1\tau_{\omega_{\mathcal{T}}} (h_{m-1})$  and $h_1= g\phi_{\mathcal{T}}(g^*)$. Given a finite group $G$ we are looking for an $m<|G|$ such that $h_m=e\Rightarrow {\omega'_{\mathcal{T}}}^m={\omega_{\mathcal{T}}}^m$. For the TC, $G=\mathbb{Z}_2=\{1,g\}$ the two phases are distinguished by $ \omega_{\mathcal{T}}=\{1,g\}$, which is gauge-invariant ($m=1$), and it follows that in both cases $\phi_{\mathcal{T}}(g^*)=g$. The phase with $ \omega_{\mathcal{T}}=g$ corresponds to a non-trivial symmetry fractionalization of the charge and it is equivalent to the braiding with the $m$ particle, resulting in a $-1$ sign. It is left for future work to understand how these phases can be distinguished for any group $G$.

%

\bibliography{Bibliografia}

\end{document}